\documentclass[12pt]{JHEP3}
\usepackage{amsmath,epsfig}
\def\be{\begin{equation}}
\def\ee{\end{equation}}
\def\ba{\begin{eqnarray}}
\def\ea{\end{eqnarray}}

\def\beqa{\begin{eqnarray}}
\def\eeqa{\end{eqnarray}}
\def\fn{\footnote}
\def\p{\partial}
\def\w{\wedge}

\def\cn{{\cal N}}

\def\d{\delta}
\def\p{\partial}

\def\ti{\tilde}
\def\oh{{1\over 2}}
\def\a{\alpha}
\def\b{\beta}

\def\sq{\sqrt{1-B\dot T^2}}
\def\cA{{\mathcal{A}}}
\title{Warped  Tachyonic Inflation in Type IIB  Flux
                Compactifications
 and the Open-String
  Completeness Conjecture}
\author{Daniel Cremades, Fernando Quevedo and Aninda Sinha \\
{\it   Department of Applied Mathematics and Theoretical
Physics,\\ Wilberforce Road, Cambridge CB3 0WA, UK} \\
\email{d.cremades,f.quevedo,a.sinha@damtp.cam.ac.uk}} \abstract{We
consider a cosmological scenario within the KKLT framework
  for moduli stabilization in string theory. The universal open string tachyon of
decaying non-BPS D-brane configurations  is proposed to drive eternal
topological
  inflation.
 Flux-induced `warping'
  can provide the small slow-roll parameters needed for
  successful inflation. Constraints on the parameter space
  leading to sufficient number of e-folds, exit from inflation,
  density perturbations and
  stabilization of the K\"ahler modulus are investigated. The
  conditions are difficult to satisfy in Klebanov-Strassler
  throats but can be satisfied in $T^3$ fibrations and other generic
                Calabi-Yau manifolds. This
  requires large volume and magnetic fluxes on the D-brane. The end
 of inflation may or may not lead to cosmic strings depending
  on the original non-BPS configuration.
 A careful investigation of initial conditions leading to a
  phenomenologically viable model for inflation is carried
  out. The initial conditions are chosen on the basis of Sen's open
  string completeness conjecture.  We find time
  symmetrical bounce solutions without initial singularities for $k=1$ FRW
  models which are correlated with an inflationary period. Singular big-bang/big-crunch
  solutions also exist but do not lead to inflation.
There is an intriguing correlation
between having an inflationary universe in 4 dimensions and 6 compact
  dimensions or
a big-crunch singularity and decompactification.} %%FF reworded some of the abstract

\preprint{DAMTP-2005-50\\  hep-th/0505252}

\begin{document}
\section{Introduction}

There has been substantial progress in deriving cosmological
inflation from string theory. Several stages of this progress can
be identified. First, the realisation that there are suitable
candidates to be the inflaton field in the low-energy spectrum of
string theory: starting with  the dilaton and geometric (closed
string) moduli \cite{bg} and later the open string modulus
describing the position of D-branes \cite{dvalitye,dbraneinflation}.

A second stage was the consideration of calculable  effective
potentials that could give rise to inflation. In this regard the
potentials for brane-antibrane systems \cite{bmnqrz,ds, bmqrz} as
well as intersecting brane systems \cite{grz, d3d7} were
considered. In these cases an elegant way to end inflation was
uncovered by means of tachyon condensation \cite{bmnqrz}. An  open
string mode becomes tachyonic at a critical brane separation and
drives the end of inflation. This field appears generically in
non-BPS D-brane systems and may be itself a candidate to drive
inflation (see for instance \cite{sa,bmnqrz,panda,calca}). However, the
absence of small parameters necessary to control the  slow-roll
conditions was a major drawback \cite{bmnqrz, linde}. This led to
the abandonment of the idea of the tachyon as the inflaton. But
its role to end inflation, realising hybrid inflation in string
theory, proved very important. Its potential  gives rise to
the possibility of a cascade of D-branes of smaller dimensionality
as topological defects, including cosmic strings
 \cite{bmnqrz}. A higher dimensional
realisation of the Kibble mechanism was also used to argue that no
domain walls nor monopoles are produced, but potentially
detectable cosmic strings are produced consistent with
experimental constraints \cite{henry}, therefore providing a
testable implication of these scenarios in terms of the remnant
cosmic strings \cite{henry}.

 The advent of semi-realistic models based on warped
compactifications with a moduli stabilisation mechanism {\it \`a
la} KKLT changed the complexion of the picture to a great extent.
The next stage of this continuous progress was to embed the
calculable brane-antibrane potentials in a set-up where all the
non-inflaton moduli fields can be stabilised \cite{kklmmt, renata,
bcsq, shamit, trivedi,berg}, following the KKLT scenario of moduli
stabilisation \cite{kklt}. So far this is the only framework in
which D-brane inflation can be claimed to be at work since in the
previous attempts the scalar potential of the geometric moduli
fields had a non-inflationary runaway behaviour. Within the KKLT
scenario, the brane-antibrane configuration can in principle give
rise to inflation in a calculable manner. One of the important
features of this scenario
 is that it
 shares the interesting consequences of the tachyon
potential mentioned previously which are very robust, including
the generic presence of cosmic strings \cite{cmp}. Furthermore,
within the KKLT scenario, an alternative to D-brane inflation was
also discovered in which the complex volume modulus is the
inflaton field, without the need to introduce brane-antibrane
configurations. This is known as  racetrack inflation {\cite{racetrack}}. It
realises  eternal topological inflation \cite{andrei}
in string
theory.

These two modern implementations of inflation represent the state of
the art within string cosmology, realising different inflationary
scenarios of the past (hybrid inflation and eternal topological inflation
respectively). They have different physical implications regarding
the density fluctuations imprinted in  the cosmic microwave
background (CMB) and from the prediction or not of remnant cosmic
strings. They also share some problems. A theory of initial
conditions is lacking, also a detailed study of reheating in a
realistic setting is yet to be achieved (for recent progress in
this direction see \cite{bbc}). Moreover, reference \cite{kklmmt}
pointed out a generic fine-tuning problem that has been known in
the context of supergravity models of inflation for some time
\cite{etaproblem}\ and
seems generic in the D-brane inflation case: the $\eta$ problem.
This refers to the fact that any field appearing in the K\"ahler
potential receives a contribution of order one to the slow-roll
parameter $\eta$, which then requires the existence of other terms
in the potential that cancel this contribution to
 an accuracy of at least $1/100$, implying an unwanted fine tuning of parameters \cite{kklmmt}.
 Actually, concrete
realisations of this fine tuning   in the brane-antibrane case
needed an slightly worse tuning
to within $1/1000$ accuracy \cite{bcsq} \footnote{Recently this
fine tuning has been
  claimed to be improved by possible cancellations in field space for
  which the parameter $\eta$ could become small \cite{henryhassan}, as well as by considering
  fast roll together with slow-roll, as in \cite{dbi}.}. A similar
tuning of
 parameters was also needed
in the racetrack scenario \cite{racetrack, graham}{\footnote{In
M-theory models there is an interesting proposal to achieve
assisted inflation without fine tuning
     \cite{bbk}. It would be interesting to complete this class of models to include volume stabilisation. }}.
%%FF corrected the footnote and added small things here and there

It is then desirable to find ways to ameliorate this fine tuning
within the KKLT scenario. For this we need to find a natural way
to generate small slow-roll parameters. Warp factors are natural
sources of small parameters in string theory. The possibility of
warped inflation was actually proposed in \cite{kklmmt} (see also
\cite{shamit}). However, this possibility was disregarded
since their inflaton field was the brane separation which is
subject to the $\eta$ problem. More recently, warping effects were
also considered in the tachyon potential
\cite{panda,calca,piao,rae} although not in a moduli-fixing
framework. It is  natural to consider the effects of warping for
the tachyon potential in the KKLT scenario which has warped
throats induced by the same fluxes responsible to stabilise the
moduli fields. This picture introduces small parameters, depending
on warping, which can alleviate the circumstances leading to the
abandonment of the tachyon driven inflation scenario. Furthermore,
the tachyon potential, originating from a non-BPS configuration,
is non-supersymmetric \footnote{Supersymmetry is actually realised
  non-linearly.}.
 It is then free from the $\eta$-problem.
This increases the possibility to obtain slow-roll inflation
without a drastic fine tuning.

On a different aspect, a general property of most inflationary
models so far is that the dynamics of inflation is very much
independent of the physics of the earlier universe. This is good
because the successes of inflation, especially regarding the
observational predictions, are not tied to the physics of the
big-bang which is largely unknown. On the other hand it would be
desirable to have completions of inflationary scenarios that
extrapolate the model to the beginning
of the universe. %%FF added the previous paragraph

In this paper we initiate a systematic search for the parameter
space in a KKLT inspired scenario with moduli stabilisation which
supports inflation with sufficient number of e-folds and satisfies
 density perturbation constraints. We will consider a simple set-up
with a single K\"ahler modulus, assuming all the other moduli to have
been stabilised, with an additional non-BPS configuration involving
the tachyon which will drive inflation. We will consider turning
  on world-volume fluxes which will aid in inflation as we will see. Schematically our model is
depicted in figure 1. In our parameter space, the
tachyon will spend a long time at the top of its potential before
rolling down and resulting in an exit from inflation. We will provide
numerical evidence that sufficient number of e-folds can be produced
in this picture. Furthermore, we will give a rough estimate for the
density perturbation and spectral index. The latter quantity can be
potentially used to falsify our model. In the example presented, the value of the
spectral index works out to be close to 0.96 and can thus be ruled in
favour or against
through experimental tests to be conducted not so distant in the
future, although values closer to 1 are also possible to obtain.

This model shares similar properties with racetrack inflation in
the sense that both give rise to eternal topological inflation.
The tachyon potential has the slight advantage of being universal
for any non-BPS configuration, therefore its implications are more
robust. On the other hand, the end of inflation is essentially the
same as  in the brane-antibrane case, because it is precisely the
tachyon field that ends inflation after reaching its minimum
value. There is a difference in the case of non-BPS branes with
respect to the brane-antibrane configurations, because in this
case the tachyon field is real and therefore cosmic strings are
not necessarily produced as topological defects since the
corresponding defects are domain walls.

%%FF took some fine tuning comment off below
An advantage over those models, refers to the initial conditions.
Tachyon condensation is supposed to produce copious amounts of
massive closed string modes which will back-react on the geometry
rendering an effective action analysis questionable \cite{mass}.
Sen's open-string conjecture \cite{ocomp,senreview,senoscomp}
allows us to work around this problem by carefully choosing the
initial conditions which allow us to trust the effective action
approach{\footnote{There is a
    recent paper \cite{singh} in which the tachyon cosmology in de
    Sitter gravity is considered. However the effect of the
    back-reaction due to the closed string modes was ignored.}}. If this conjecture
holds,
as  several arguments  suggest \cite{senreview},
 then we can not only use the effective field theory to
describe inflation but can actually use it to investigate the time before
inflation, including the big-bang and before. It is remarkable that we find
bounce solutions without initial singularity with a pre big-bang period
identical to the post big-bang due to the time symmetry of the
solutions.
This differs from the original pre big-bang scenario \cite{veneziano}
in the sense that
in that case the Hubble parameter tends to blow-up at $t=0$ whereas from
our initial conditions $H=0$ at the origin.
We also find that the existence of an inflationary period naturally
correlates with the bounce solutions whereas singular solutions do not
lead to inflation.

Our paper is organised as follows. In section 2, we summarise  the
 realisation
 inflation using the tachyon field as the inflaton. In section 3, we
review Sen's open string completeness conjecture. In section 4, we
dimensionally reduce the 10-d action in a warped set-up and add the
dimensionally reduced non-BPS system of branes. The K\"ahler modulus
stabilisation is achieved through the KKLT mechanism. In section 5, we
study inflation in this set-up. Sections 7 and 8 are devoted to
 concrete examples where the effects of warping on the metric as well
 as the fluxes can be estimated. In section 9, numerical evidence is
presented supporting our scenario as a viable candidate for an
inflationary model. We conclude with a discussion and open questions
in section 10.
\begin{figure}
\begin{center}
\epsfig{file=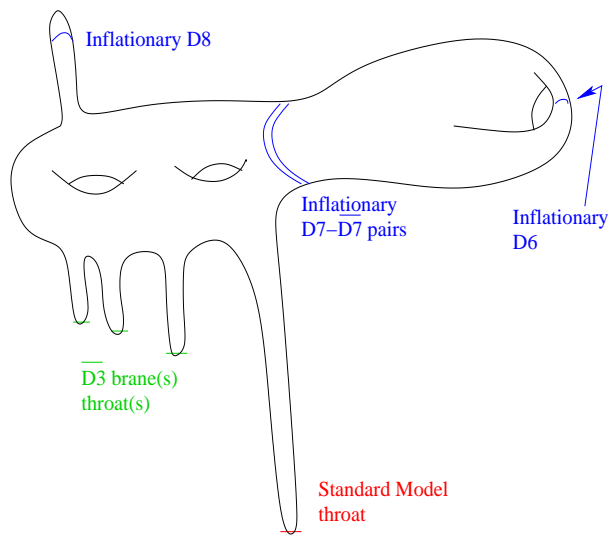,width=8cm,height=7cm}
\caption{The general setup for our model is IIB string theory compactified in a warped CY as shown here.
The warped CY has a complicated structure with several throats in which the warping is big and several other parts where the warping
is milder. The SM is placed generically in one of the throats, but its precise embedding is not specified (also, the fact that it is on a throat is
not crucial for the model to work). There are a series of anti D3s placed on throats as in the KKLT setup. Inflation is produced by one or several
non-BPS systems like D6s wrapping toroidal 3-cycles, D7s-anti D7s on top of each other wrapping general cycles or D8s in the middle of KS throats.
Presence of electromagnetic flux on these systems is generic. }
\end{center}
\end{figure}

\section{Tachyon Inflation}
In this section we will review the basic material needed to
describe tachyon inflation in the presence of warping. We will
start collecting some basic equations which will be needed later
on.

\subsection{Inflation Basics}

In an effective field theory in 4d,  slow-roll inflation is
obtained if a scalar potential, $V(\phi)$, is positive in a region
where the slow-roll conditions are satisfied:
\begin{equation}
\label{slowroll} \epsilon  \equiv  \frac{M_{pl}^2}{2}\ \left(
\frac{V'}{V}\right)^2\ \ll  \ 1\ , \ \ \ \  \qquad \eta  \equiv
M_{pl}^2\ \frac{V''}{V}\  \ll  \ 1\ .
\end{equation}
Where $M_{pl}$ is the rationalised Planck mass ($(8\pi G)^{-1/2}$)
and primes refer to derivatives with respect to the scalar field,
which is assumed to be canonically normalised. In the region where
the slow-roll conditions are satisfied the scale factor suffers an
exponential expansion $a\sim e^{Ht}$ with the number of e-foldings
 given by:

\begin{equation}
 N(t)\equiv \int_{t_{init}}^{t_{end}} H(t') dt'\  =\
\frac{1}{M_{pl}^2} \int_{\phi_{end}}^{\phi_{init}}
\frac{V}{V'}\,  d\phi\ . \end{equation}

A successful period of inflation required to solve the horizon
problem usually needs at least $N\geq 60$. Slightly smaller values are allowed
depending on the inflation scale.

The amplitude of density perturbations is given by :
\be\label{deltaH} \delta_H = \frac{2}{5} {\mathcal P}_{\mathcal
R}^{1/2} =
   \frac{1}{5 \pi \sqrt{3}}\,{V^{3/2}\over M_{pl}^3\,V'}= 1.91\times 10^{-5}\,,
\ee
where ${\mathcal P}_{\mathcal R}$ is the power spectrum computed
in terms of the two-point correlators of the perturbations. The
numerical value is given by the COBE normalisation which must be
computed at the point where  the last $60$ e-folds of inflation start. The
spectral index is given by
\be\label{sindex} n-1 = \frac{\partial\ln{\mathcal P}_{\mathcal
R}}{\partial\ln k }
       \simeq 2\eta - 6\epsilon  \,,\qquad\qquad
{dn\over d\ln k} \simeq 24 \epsilon^2 - 16\epsilon\eta +2\xi^2 \,.
\ee
where $k$ is the length scale
and  $\xi^2\equiv M^2_{pl} \frac{V'\,V'''}{V^2}$. This shows  that for
slow rolling ($\eta,\epsilon \ll 1$) the spectrum is almost scale
invariant ($n\sim 1$).

\subsection{Topological Inflation}

Satisfying the slow-roll  conditions is not an easy challenge for
typical potentials since the inflationary region has to be very
flat. Furthermore, after finding such a region we are usually
faced with the issue of initial conditions, why should the field
$\phi$ should start in the particular slow-roll domain.

Topological inflation was introduced in \cite{andrei} partly
as a way to ameliorate the issue of initial conditions of standard
inflationary models. The idea is that usually starting the slow
roll close to a maximum of a one-field potential is convenient
because the corresponding parameter $\epsilon$ is naturally small
there, leaving only the problem of finding a small enough $\eta$,
which will be small enough if the maximum is very flat. But in
general starting very close to this maximum is a very fine tuned
configuration.

In topological inflation the initial conditions are governed by
the existence of a topological defect at the maximum of a
potential with degenerate minima. If the defect is thicker than
the Hubble size $H^{-1}$, then a patch within the defect  can
inflate in all directions (including the ones transverse to the
defect) and become our universe. The conditions for the thickness
of the defect happen to be satisfied if the slow-roll condition,
requiring $\eta$ to be small, holds. This is easy to see since the
thickness of the defect ($\delta$) is inversely proportional to
the curvature of the potential at the maximum which itself is
proportional to $\eta$. Then,  the condition $\delta\gg H^{-1}$ is
satisfied if $\eta \ll 1$.

Topological inflation never ends, but the region that becomes our
universe does so by standard rolling of the potential towards one
of the minima, ending inflation once the slow-roll conditions
cease to hold.
 Therefore this is a particular case of the  standard slow-roll inflation and the
topological defect only provides the natural initial conditions to
start rolling close to the maximum. As such all the implications
of small field inflation apply equally well here, including the
spectrum of density perturbations. Notice that several
inflationary potentials give rise to eternal inflation either due
to quantum fluctuations or classical effects like in topological
inflation (for a recent discussion of eternal inflation see
\cite{gt}).

\subsection{Tachyon Inflation}

In a model with tachyon driven inflation, the starting point is a
4-dimensional effective action given by
\be
S= \int \,d^4x\sqrt{-g}\left({M_{pl}^2\over 2} R-\cA
V(T)\sqrt{1+B\partial_\mu T\partial^\mu T}\right)+{\rm CS}\,,\label{actioneg1}
\ee
where $M_{pl}$ is the 4-d Planck-mass, $\cA$ and $B$ are positive
dimensionless parameters which come from dimensionally reducing the effective action
for an unstable probe brane system and usually depend on the metric
warp factors,
the string coupling $g_s$, string length $l_s=2\pi\sqrt{\alpha'}$, the electric and magnetic flux that can be
turned on in the world volume of the brane
and the volume of the compact manifold. CS
stands for the terms coming from the Chern-Simons part of the brane
action. $V(T)$ is the tachyon potential which string field theory
analysis suggests should be taken to be
\be \label{coshpot}
V(T)={V_0\over \cosh({T\over \sqrt{2\alpha'}})}\,.
\ee
Where $V_0$ is a dimension 4 parameter (in units of the string scale
$m_s$) and it is proportional to $1/g_s$ and the volume of the cycle
that the corresponding non-BPS brane is wrapping \footnote{Notice that $V_0$
can be absorbed in the definition of the parameter $\cA$, as we will do
in the next sections. Here we will keep it to make explicit the
different sources of scales.}.

One
considers a homogeneous tachyon field $T(x^0)$ which acts as a perfect
fluid source for gravity.
Notice that the tachyon field $T$ in action (\ref{actioneg1}) does not
have canonical kinetic terms. We can define a canonically normalised
field $\phi$ such that:
\be
\phi \ =\ \sqrt{\cA B}\int \sqrt{V(T)}\ dT
\label{canon}
\ee
It is clear that expanding the square root in
(\ref{actioneg1}), the field $\phi$ will have canonical kinetic
terms. It is with respect to $\phi$ that the slow-roll conditions
apply.
Using this relation between $\phi$ and $T$ we can rewrite the
slow-roll conditions in terms of the field $T$ as follows:

\begin{equation}
\label{slowroll2} \epsilon  =  \frac{M_{pl}^2}{2\cA B}\ \left(
\frac{V'^2}{V^3}\right)\ \ll  \ 1\ , \ \ \ \  \qquad \eta  =
\frac{M^2_{pl}}{\cA B}\ \left[\frac{V''}{V^2}\ -\ \frac{1}{2}
  \left(\frac{V'^2}{V^3}\right) \right] \ll  \ 1\ .
\end{equation}
Prime now denotes
differentiation with respect to $T$.
Similarly, the number of e-foldings takes the form \cite{rae,sv}:
%Ignoring the Chern-Simons piece, it was
%shown in \cite{rae,sv} that in the slow-roll approximation and for a
%spatially flat universe, the number
%of e-folds  of inflation is given by
%
\be
N(T)\approx \frac{\cA B}{M_{pl}^2}\  \int_{T_e}^{T_b} {V^2\over V'}\,  dT\,,
\ee
where $T_b$ is the tachyon field value at the beginning of inflation
and $T_e$ is the value at the end of inflation.
The density perturbations take the following form in terms of the field $T$.
\be
\delta_H\ =\ \frac{\sqrt{\cA^2 B}}{5\pi \sqrt{3} M_{pl}^3}\,
  \frac{V^2}{V'}.
\ee

In the absence  of warping, as argued in \cite{linde}, in order to get
many e-folds, one needs either a large string coupling or small
compactification volume. In both cases, $\alpha'$ corrections become
very important and render the lowest order analysis incomplete and
possibly wrong. Furthermore, it is clear that the tachyon potential
does not have small parameters and cannot provide small values of
$\epsilon$ and $\eta$, nor large values of $N$.
The presence of the parameters ${\cal A}$ and $B$  makes the situation much
better and amenable to analysis using the effective theory as argued
in \cite{rae}.

We can explicitly compute the slow roll parameters for the potential
$V(T)$ in equation (\ref{coshpot}). This gives:
\be
\epsilon = \frac{M_{pl}^2}{4\alpha' \cA BV_0}\
\frac{\sinh^2 x}{\cosh x},
\qquad \eta=  -\frac{M_{pl}^2}{2\alpha' \cA BV_0}\
\cosh x\left[1-\frac{3}{2}\,  {\rm {tanh}}^2 x\right]
\ee
With $x\equiv \left(\frac{T}{\sqrt{2\alpha'}}\right)$.
It is now clear that for large enough values of $\cA B$ the slow-roll
conditions can be easily satisfied, giving rise to successful
inflation. Notice that the scalar potential in terms of the
canonically normalised field $\phi$ has a maximum at the origin with
two minima at values of order $\sqrt{\cA B}$ and therefore satisfies the
conditions for topological inflation.

The number of e-foldings can be computed to be:
\be
N\ \approx -\frac{2\alpha' \cA BV_0}{M_{pl}^2} \
\log\left(\frac{T_b}{\sqrt{8\alpha'}}\right)
\ee
 which can easily satisfy $N\geq 60$ for large enough warping
 $\cA B$. The minus sign suggests that for the formula to be valid,
 $T_b<\sqrt{8 \alpha'}$.
The density fluctuations can also be estimated to be
\be
\delta_H = \sqrt{\frac{\alpha'\cA^2 V_0^2 B}{75\pi^2 M_{pl}^6}}%\over M_{pl}^3}
\, { \rm csch }\, \left(\frac{T_b}{\sqrt{2\alpha'}}\right) \ee
Choosing $T_b$ to correspond to the start of the last $60$
e-folds, we can easily satisfy the requirements $N=60$, $\delta_H=
1.92 \times 10^{-5}$ by adjusting three dimensionless parameters,
namely $\cA$, $B$ and the ratio $\gamma\equiv m_s/M_{pl}$ which
itself depends on the volume of the extra dimensions and warp
factors that will be explicitly defined in later sections. To get
enough e-foldings (and satisfy slow-roll conditions) we need to
have  $\cA B \gamma^2\sim 10^2$ and
to satisfy the COBE normalisation % $\delta_H= 1.92 \times
%10^{-5}$
we need $\cA^2 B \gamma^6 \sim 10^{-10}$. These conditions can be
satisfied very easily by properly choosing brane configurations,
warped throats and moduli stabilisation at sufficiently large
volume (in units of the string length). For instance  $\cA\sim 1$,
$B\sim 10^6$ and $\gamma \sim 10^{-3}$ satisfy all the
requirements, as well as $\cA^{-2}\sim B\sim 10^8$, $\gamma\sim
\left({\cal O}(10^{-2})- {\cal{O}}(10^{-1})\right)$. Notice that
the %warp-depending 
factors ${\cal A}$, $B$ cannot be simultaneously large
to satisfy both conditions.  Therefore not every warped brane
configuration will give rise to successful inflation
\footnote{More freedom
  may be obtained if this can be incorporated in a curvaton scenario
  in which both constraints are independent. For a recent discussion
  see \cite{lyth}.}. But some of them will.
%$AB\sim 10$ with $A\sim 10^{-9}$ and the string scale
%$m_s= \left({\cal O}(10^{-2})- {\cal O}(10^{-1})\right)  M_{pl}$. Since
%$A,B$ generically involve the warp factor, these numbers are quite natural.

This is very encouraging. However, one cannot choose the
parameters $\cA$ and $B$ arbitrarily as one needs to take into
account the stabilisation of the volume of the compact manifold.
In a realistic string model, the tachyon potential will depend on
the dilaton and geometric moduli, which in the absence of other
sources of potential, will have a runaway behaviour. Therefore, it
is not consistent to concentrate only on the tachyon field
dependence of the scalar potential in order to look for
inflationary trajectories. Tachyon inflation should therefore be
considered within a mechanism that fixes the moduli. Furthermore
the choice of parameters has to be consistent with the effective
field theory description in which the Kaluza-Klein masses have to
be suppressed with respect to the string scale but larger than the
Hubble scale. This will place strong restrictions on the choice as
we shall see in the next sections, where we will describe these
issues explicitly in the KKLT scenario of moduli stabilisation.

\section{The Open-String Completeness Conjecture}

One important observation at this stage is as follows\fn{To the best of our 
knowledge, the only other paper in which Sen's open-string completeness
conjecture has been applied to cosmology before is \cite{Ghodsi}.}. It has been
known for a while \cite{senreview,mass,ocomp} that the decay of an unstable
brane will produce very massive closed string modes with masses
comparable to $1/g_s$. These massive modes will back-react on the
geometry potentially rendering the effective action analysis
invalid. In order to take into account this back reaction, Sen
has suggested an open-string completeness conjecture\cite{senreview,ocomp}
which we will review now.

According to the open-string completeness conjecture there is a quantum open string field theory that describes
  the full dynamics of an unstable D-brane without an explicit
  coupling to all the vibrational modes of the closed string. The classical results are supposed to
  correctly describe the evolution of the quantum expectation values
  in the weakly coupled open string field theory. The open string
  theory describes a consistent subsector of the full string theory
  and is capable of capturing the quantum decay process of an unstable
  D-brane. Although this
  conjecture has never been proved, there is strong evidence that it
  holds \cite{senreview}. We will assume that this conjecture holds for the rest of the
  paper. This conjecture places constraints on the initial conditions
  that one can impose on the closed string fields in the effective
  action, for example the graviton. To demonstrate this let us review
  Sen's analysis in \cite{senoscomp}. There the starting point was the
  action in (\ref{actioneg1}) with an additional cosmological constant
  term and with $\cA=A_0, B=1, M_{pl}=1$. For our analysis we will ignore this
  cosmological constant term as it will play no role to demonstrate
  the constraints following from the open-string completeness
  conjecture. Let us assume a spatially homogeneous time dependent
  solution in the FRW form:
\ba
T&=& T(x^0) \nonumber \\
ds^2 &=& -(dx^0)^2+a(x^0)^2 \left({dr^2\over
  1-kr^2}+r^2d\theta^2+\sin^2\theta d\phi^2\right)\,, k=1\,. \label{FRW}
\ea
The equations of motion for $a$ and $T$ that follow from
  (\ref{actioneg1}) are
\ba
{\ddot a \over a}&=& {A_0 V(T)\over 6 \sqrt{1-\dot T^2}}\left(1-{3\over 2}
\dot T^2 \right)\,,\\
3H {V(T) \dot T\over \sqrt {1-\dot T^2}}+\partial_t\left({V(T) \dot
  T\over \sqrt {1-\dot T^2}}\right)&=&-\sqrt{1-\dot T^2} V'(T) \label{eoms1}
\ea
where $H=\dot a/ a$ with the Friedman constraint
\be
H^2={1\over 6} {A_0 V(T)\over \sqrt{1-\dot T^2}}-{1\over a^2}\,.
\ee
Once the Friedman constraint is satisfied at an initial time, it is
  satisfied at all times since its time derivative leads to the
  equations of motion (\ref{eoms1}).
For the two second order differential equations, we need four
initial conditions. The Friedman equation imposes one constraint.
One can choose the origin of time by imposing $\dot T=0$ at
$x^0=0$ thereby eliminating an additional condition. Thus we are
left with two possible initial conditions. Naively one would take
these to be $T(x^0)=T_0$ and $\dot a(x^0)=a_0$. The open string
completeness conjecture reduces this two-parameter space of
solutions to a one-parameter family of solutions. This is done by
noting that the rolling tachyon solution describing the decay of a
D-brane in open string theory is time reversal symmetric. Thus the
solutions following from the above equations of motion must also
be time reversal symmetric. For this to be possible, we must
choose $a(x_0)=a(-x_0)$ thereby imposing $\dot a=0$ at $x_0=0$.
Note that this
  implies that solutions with a big-bang singularity will also have a big-crunch. Sen found time
reversal symmetric solutions in \cite{senoscomp}, however he found
that the slow-roll conditions could not be met and hence
sufficient number of e-folds could not be obtained. We will
perform a more detailed analysis motivated by this in a KKLT
inspired scenario and include the effects of warping which would
impose $B\neq 1$ and also introduce warp factor dependence in $A$.
In our simple model we will also include a K\"ahler moduli to
study what happens to the compact manifold during this period of
cosmological evolution of our
  universe. We will find that in our models getting sufficient number
  of e-folds excludes the scenario of
  big-bang-big-crunch. {\footnote{We would like to stress that
  time-reversal symmetric solutions are not necessarily the only ones
  allowed but rather the ones that can be described by the effective
  action that we are considering. It is probably possible to describe
  time-asymmetric solutions with more complicated configurations of
  D-branes whose description is not captured by this effective action.}}

\section{Non-BPS D-branes  and Flux Compactification}

\subsection{Dimensional reduction of flux compactification}
Type IIB  string compactifications on Calabi-Yau orientifolds with
three-form fluxes have a flux-induced warped metric \cite{GKP} .
It can be written as:

\be ds^2=e^{2\omega-6u+6u_0}\bar g_{\mu\nu}dx^\mu dx^\nu +
e^{-2\omega+2u} \ti g_{mn} dy^mdy^n \label{metric_gral}\ee

 where the 4d spacetime coordinates are $x^\mu$, $\mu=0,1,2,3$ whereas the internal space coordinates
 are $y^m$ $ m=4, \cdots 9$. Here $\omega(y)$ is a
function of the internal coordinates of the manifold that  defines
the warp factor, and $u(x)$ is a four-dimensional field
corresponding to the real part of the K\"ahler modulus that
determines the volume of the compact space. We will suppose that
the internal metric has only one K\"ahler modulus (whose imaginary part
is $e^{4u(x)}$) and a number of complex structure moduli. $u_0$ is a
dimensionless constant whose magnitude will be fixed afterwards.
After dimensional reduction, all the complex structure moduli and
the dilaton get fixed to its value at some minimum of the
potential derived from the GVW superpotential \cite{GVW,GKP}
generated by the flux of the complex type IIB form $G_3$. We will
consider all the excitations of these complex structure moduli
fields to have been consistently integrated out from four
dimensional physics. The four dimensional action reads then
\cite{Buchel}

\beqa
S={M_{pl}^2\over 2}\int d^4x \sqrt{-\bar{g}} \left[\bar R -24
  (\bar\nabla u)^2 \right]\label{action_4D}
\eeqa

with

\beqa
{M_{pl}^2\over 2}={e^{6u_0}\over (2\pi)^7g_s^2 \a^{\prime 4}}\int d^6y
 \sqrt{-\bar g} ~e^{-4\omega(y)}.
\eeqa

Being a supersymmetric construction, this action can be put in
terms of a K\"ahler potential and a superpotential. If we define
the K\"ahler modulus to be $\rho(x)\equiv Y(x)+iX(x)$, with
$X(x)\equiv e^{4u(x)}$, and the axion field $Y$ is defined from
the RR 4-form $C_{(4)}$. Then the four dimensional action
(\ref{action_4D}) becomes (where an appropriate kinetic term for
$Y$ has been added)

\beqa
S={M_{pl}^2\over 2}\int d^4x \sqrt{-\bar{g}} \left(\bar R -6{\bar\nabla_\mu \bar \rho \bar \nabla^\mu \rho\over |\rho-\bar \rho|^2}\right)
\eeqa
that has the canonical form for a $\cn=1$ supergravity action with
K\"ahler potential $K=-3~{\rm log} [-i(\rho-\bar \rho)]$.
From now on, we will assume the field $Y(x)$ to be
stabilised at some minimum (we will be more concrete in the following sections) and suppose its fluctuations to decouple from the
spectrum at low energies.

The vev of $X$ is related to the value of the volume of the compact dimensions. This volume is given by

\beqa
{\rm Vol}_6=\int d^6y\sqrt{g_6}=\langle X\rangle^{3/2}\int d^6y\sqrt{\ti g}~e^{-6\omega}
\eeqa

\subsection{Dimensional reduction of the tachyon action}

Consider the following (string frame) action for the non-BPS D$p$-brane in Type IIB
string theory ($p$ even)\fn{We consider
$\p_yT=0$ so that the CS term is zero. This is consistent with the fact that in absence of a magnetic field the
zero mode of $T$ will be a constant over the compact directions.}. This
action is also valid for a $D\bar D$ system where the branes are
sitting on top of each other and with no world-volume gauge field
turned on, in which case one has to change $T_p\to \sqrt{2} T_p$.

\beqa
S_{T}=-T_p\int_{M_4\times \Sigma} d^{p+1}x \,
V(T) \sqrt{|g_{ab}+\p_a T\p_bT|}
\label{action_tach}
\eeqa

The change from the string frame to the Einstein frame is given by a
change in the metric of the form
$g_{\mu\nu,E}=e^{\phi_0-\phi}g_{\mu\nu,s}$, where $\phi_0$ is the
vev of the dilaton \cite{polch2}. In the Einstein frame the tension
of the Dp brane reads

\beqa T_p={\sqrt{2}(2\pi)^{-p}\a'^{-{{p+1}\over 2}}\over g_s}. \eeqa

For $T(x)$ only depending on time
%\beqa
%\sqrt{|g_{ab}+\p_a T\p_bT|}=\sqrt{|g_4|}\sqrt{|g_{int}|}\sqrt{1-|g^{00}|\dot T^%2}\,.
%\label{skrt}
%\eeqa
and assuming that the warping is only transverse to the
brane{\footnote{There is in principle an additional term dependent on
    the internal coordinates in the action
involving $B_{ab}$ \cite{lyth} but since this is only turned on in the compact
    directions, it will be exponentially suppressed, noting that the
    compact metric $g^{mn}$ is proportional to $e^{2\omega}$. However
    a constant $B_{ab}$ can actually be useful to get inflation as we
    will explain later.}}  , we finally find

\beqa
S_{T,DBI}&=&-\sqrt{2}(2\pi)^{-p}\a'^{-{p+1 \over 2}}g_s^{-1}h_{p-3}\ti{\cal V}^2\times \nonumber\\
&\times&\int d^4x\sqrt{|\bar g|}
e^{(p-15)u}e^{(7-p)\omega (y_p)}V(T)\sqrt{1-e^{-2\omega+6u-6u_0}\, \dot T^2}\nonumber\\
\eeqa
where $\omega (y_p)$ stands for the value of $\omega (y)$ in the position of the brane.
%The $g_s^{-1/2}$
%factor arises when changing to the Einstein frame.
We have assumed $|\bar g^{00}_E|=1$, that is what we are going to find hereafter.
We have moreover defined

\beqa h_{p-3}\equiv \int d^{p-3}y\sqrt{|\ti g_{int}|}. \eeqa This
factor will depend on the Calabi-Yau under consideration; we will see examples below of its value in several
scenarios. We can
consistently add this action to (\ref{action_4D}) provided that the
addition of this term can be seen as the addition of a D-term. This
can be argued by noting that the 32 bulk supersymmetries are
non-linearly realised in the world-volume for the non-BPS branes.

%\subsection{The case for a D9-D9bar system}

%We can consider the case of a D9-D9bar system in which just the tachyon (and not% a Wilson line) produces inflation. The action is
%still the same (\ref{action_tach}), but in this case the DDbar system occupies t%he whole space and we cannot consider warping along
%the transverse directions since there are no transverse directions. Alternativel%y, a warping equal to 1 would give no inflation,
%see below, and a constant warping would be useless since it could
%always be reabsorbed in $u(x)$. We must consider another strategy.

%Let us start from equation (\ref{skrt}). We have

%\beqa
%\sqrt{|g_{ab}+\p_a T\p_bT|}&=&\sqrt{|g_4|}\sqrt{|g_6|}\sqrt{1-b|g^{00}|\dot T^2}%\\
%&=&\sqrt{|\bar g|}\sqrt{|\ti g_6|}~e^{-2A(y)-6u(x)}\sqrt{1-e^{-2A(y)+6u(x)}\dot %T^2}\nonumber
%\eeqa

%In the following we will factor out of the whole action (gravity + tachyon) a fa%ctor of $M_{Pl}^2/2$. This will imply that we will
%consider terms of the form

%\beqa
%{\int d^{10}x }
%\eeqa

\subsection{The 4D potential for the K\"ahler modulus}

We can add a potential term for $u$ following KKLT, but we must argue
that these non-perturbative corrections will not
mix with the tachyon. A heuristic argument for this is as
follows. Non-perturbative corrections arise from evaluating the
partition function at Euclidean saddle points. For a finite
contribution, one needs a finite action for an Euclidean object which
then gets exponentiated. In the KKLT set up the Euclidean object under
consideration is a D3 brane which wraps a 4-cycle of the compact
manifold. If one added an anti-D3 brane, it could annihilate this D3
brane if it wrapped the same 4-cycle as a result of which the
non-perturbative potential would need to get modified by a function
that depended on the position of the anti-D3 brane. Such a
complication will not arise in the cases that we will consider. In
principle, however, euclidean non-BPS D$p$ branes could wrap some cycle of
the compact manifold and produce a finite action which would add to
the non-perturbative superpotential. Such contribution would look like
$$
C(T)\,  e^{-b(T) X^{(p+1)/4}}\,,
$$
with $C(T)\rightarrow 0$ as the tachyon condenses. Since $p+1>6$ in
the cases where we get inflation, such terms will not contribute.
Furthermore we envision the non-BPS brane generating the tachyon
potential
to be separated from the four-cycles where the gauge theory generating
the non perturbative superpotential sits.

The four dimensional action for a single chiral multiplet $\rho$
is given by\fn{We follow the conventions of \cite{DouglasKachru}.}

\beqa
S={M_{pl}^2\over 2} \int d^4x \sqrt{-\bar g} \left(\bar R -g_{\rho\bar\rho}\bar\nabla_\mu \rho \bar\nabla^\mu\bar\rho- V_F(\rho)
\right)+({\rm D-terms})
\eeqa
with $g_{\rho\bar\rho}=\p_\rho\p_{\bar\rho}K$, $g^{\rho\bar\rho}=(g_{\rho\bar\rho})^{-1}$
and $D_\rho W=\p_\rho W+W\p_\rho K$, and

\be
V_F(\rho)={M_{pl}^2\over 2\pi}
e^K\left [g^{\rho\bar\rho}D_\rho W \overline{D_{\rho}W} -3|W|^2\right]
\ee
is the standard F-term part of the supergravity scalar potential.
Let us then assume a KKLT-like superpotential for $\rho$ (remember we are going to use $e^{4u}={\rm Im}\rho$)

\beqa
W=W_0+De^{ic\rho}
\eeqa
The field $Y$ in $\rho$ can be easily stabilized
 at $\langle Y \rangle =\pi$ and integrated out, so that
$\rho=\pi + iX$. This yields\fn{Note the different sign in $W_0$ with respect to
\cite{kklt}. It arises because of the fact that the field $Y={\rm Re~}\rho$ stabilizes at $\langle Y \rangle = \pi$ instead of
$\langle Y \rangle = 0$ as assumed in \cite{kklt}, see \cite{racetrack} for details. However, using the opposite value for $W_0$
than in \cite{kklt} we can still use their same numerology.}
, using $K=-3{\rm log}(-i(\rho-\bar\rho))$

\beqa
V_F={M_{pl}^2\over 2\pi}{cDe^{-cX}\over 2X^2}\left( {1\over 3}\,
cDXe^{-cX}-W_0+D\, e^{-cX}\right)
\label{pot_KKLT}
\eeqa
This potential has a supersymmetric minimum.
%\beqa
%W_0=-De^{-cX_{crit}}-{2\over 3}DcX_{crit}e^{-cX_{crit}}.
%\eeqa
%This minimum is found doing $D_\rho W=0$. The potential for this minimum is
%\beqa
%V_{AdS}=-{M_{pl}^2\over 2\pi}{c^2D^2e^{-cX_{crit}}\over 6X_{crit}}
%\eeqa
We can add, following KKLT, a set of anti D3-branes breaking supersymmetry in the form of a D-term in another throat.
The action for an anti-D3 brane is
\be
S_{\overline{D3}}=-T_3\int d^4x \sqrt{|g_4|}=T_3\int d^4x \sqrt{|\bar g|}e^{4 \omega(y_i)} e^{-12u}=
T_3\int d^4x \sqrt{|\bar g|}e^{4 \omega(y_i)} X^{-3}
\ee
$y_i$ specifies the position of the anti D-brane $i$ in the CY.
These anti-D3s are attracted to the tip of the throat
and prefer to sit there. There  $e^{4 \omega(y_i)}\sim X~{\rm exp}(-{8\pi K\over 3 g_s M})$,
where $M$ and $K$ specify the RR and NSNS fluxes
turned on in the throat. The whole effect is to add the potential (\ref{pot_KKLT}) a term
like

\beqa
\d V={E\over X^2}
\eeqa
with

\beqa E={2\over M_{pl}^2}T_3\sum_i e^{-{8\pi K_i\over 3 g_s
M_i}}=(2\pi)^{-3} \a^{\prime-2}g_s^{-1}
\sum_i e^{-{8\pi K_i\over 3 g_s M_i}} \eeqa
Note that the potential
$V_{\rm total}=V+\delta V$ scales as $\mu^3 V_{\rm total}$ under
\be
D\rightarrow \mu^{3/2} D,\quad W_0\rightarrow \mu^{3/2} W_0, \quad
E\rightarrow \mu^3 E\,, \label{rescaling} \ee
keeping the minimum
unchanged. Other interesting scaling property of $V_{\rm total}$ is that under the change
\be
D\rightarrow \lambda^{3/2} D,\quad W_0\rightarrow \lambda^{3/2} W_0, \quad
E\rightarrow \lambda^2 E, \quad c\rightarrow \lambda^{-1}c \,, \label{rescaling2} \ee
the potential is invariant provided that $X\to\lambda X$, that is, the whole structure of extrema gets shifted
horizontally.
One can investigate a wider parameter space using these
scaling properties. Furthermore, for bigger $W_0$ one should take into
account the $\alpha'$ corrections. We will put in the first
$\alpha'$ correction that comes from the $R^4$ term in IIB
\cite{bbhl, bb, bbcq, CQS}. This is given by \be
V_{\alpha'^3}={9\sqrt{5}\xi \mu^3 W_0^2\over 4\sqrt{2} X^{9/2}}\,,
\ee where $\xi=-\chi(M)\zeta(3)\alpha^{\prime 3}/(16 \pi^3)$,
$\chi(M)$ being the Euler number of the compact manifold. We choose
$\xi \approx 1$ although one can easily accommodate other choices
allowing for larger $\xi$.

\subsection{Summarising: The total 4D action}

Putting all these pieces together we find (we will assume that $u$ only depends on the time coordinate)

\beqa
S&=&{M_{pl}^2\over 2}\int d^4x\sqrt{-\bar g}\left(\bar R_4 +24 (\dot u)^2-\hat{V}(u)\right)\nonumber\\
&&-(2\pi)^{-p}\a'^{-{p+1 \over 2}}g_s^{{p-3 \over 4}}h_{p-3}e^{(15-p)u_0}\times \label{total_action}\\
&&\times\int d^4x\sqrt{|\bar g|}
e^{(p-15)u}e^{(7-p)\omega(y_p)}V(T)\sqrt{1-e^{-2\omega+6u-6u_0}\dot T^2}\nonumber
\eeqa

with

\beqa\label{sombrero}
{\hat {V}}(X) &\equiv &V_F +\d V + V_{\alpha'^3}\nonumber \\
&= &{\mu^3 M_{pl}^2\over 2\pi}{cDe^{-cX}\over 2X^2}\left({1\over 3} cDXe^{-cX}-W_0+De^{-cX}\right)+{E\over X^2}+V_{\alpha'^3}\,,
\eeqa
where we have explicitly put in the scaling parameter $\mu$ and $X\equiv e^{4u}$.
Notice that the full potential as a function of $X$ and $T$ can be
written (in the small kinetic energy approximation for the tachyon)
as:
\be
V_{total}(X,T) \ =\ \hat{V}(X)\ + V_{tachyon}\,.
\ee

\section{Four dimensional effective action and inflation}

\subsection{Friedman equation}

Consider the following 4D action{\footnote{Note that ${M_{pl}^2\over
      2} A=\cA$.}}
(we will later substitute the actual values of the parameters but we will leave them
as parameters in order not to lose generality:

\beqa
S={M_{pl}^2\over 2}\int d^4x\sqrt{ -g}\left\{R -2\Lambda -\b(\p u)^2
-\hat V(u)-
A V(T)\sqrt{1+B(\p T)^2}\right\}
\label{eff_action}
\eeqa
With the standard FRW ansatz for the metric (\ref{FRW})
we get the following equations of motion for the metric, assuming a
homogeneous ansatz $T(x)=T(x^0)$ and $u(x)=u(x^0)$,

\ba
H^2&=&{1\over 3 M_{pl}^2}\rho-{k\over a^2}\,,
\label{friedmann}\\
{\ddot a\over a}&=&-{1\over 6 M_{pl}^2}\left(3p+{\rho}\right)\,,
\ea
where
\beqa
\rho&=&{M_{pl}^2}\left\{\Lambda+\oh \b\dot u^2+\oh \hat{V}(u) +\oh {A \cdot V(T) \over \sqrt{1-B\dot T^2}}\right\}\,,\\
p&=&{M_{pl}^2}\left\{-\Lambda+\oh \b \dot u^2 -\oh \hat{V}(u)-\oh A\cdot V(T)\sqrt{1-B\dot T^2}\right\}\,.
\eeqa

\subsection{Summary, values for the parameters and initial conditions}

Including the equations of motion for  $X\equiv e^{4u}$ and $T(t)$, we
find the following system of equations:

\beqa
3H{AVB\dot T\over \sq}+\p_t\left({ABV\dot T\over \sq} \right)&=&-A\sq ~V'(T)
\label{tachy}
\eeqa
\beqa
\b \p_t\left({\dot X\over 4X}\right)+{3\b H\dot X\over 4X}+2X\p_X\hat{V}(X)+2X\p_X\left(AV(T)\sq\right)=0
\label{closed_modulus}
\eeqa
\beqa
{\ddot a\over a}=-{1\over 3}\b \left(\dot X\over 4X\right)^2 +{1\over
  6}\hat{V}(X) +
{A \cdot V(T)\over6 \sqrt{1-B\dot T^2}}
\left({1-{3\over 2}B\dot T^2}\right)
\label{evolutiona}
\eeqa
with
$\hat{V}(X)$ given in equation (\ref{sombrero}).
%={\mu^3 M_{pl}^2\over 2\pi}\left[{cDe^{-cX}\over 2X^2}\left({1\over
%3}
%cDXe^{-cX}-W_0+De^{-cX}\right)\right]+{E\over X^2}+W_{\alpha'^3}
%\eeqa
Comparing (\ref{total_action}) to (\ref{eff_action}), the values of
the
parameters are
%\fn{The term in $A$ as $\a^{-\prime 2}$ arises
%because of a cancelation between $h_{p-3}=\a^{\prime (p-3)/2}$
%and the $\a '$ term arising from $T_p$.}

\beqa
\b&=&24\\
A&=&{2\sqrt{2}\over M_{pl}^2}(2\pi)^{-p}\a'^{-{(p+1)\over 2}}g_s^{-1}
e^{(7-p)\omega(y_p)}X^{(p-15)\over 4}\ti{\cal V}^2h_{p-3}\label{parameter_A}\\
B&=&{e^{-2\omega(y_p)}X^{3/2}\over \ti{\cal V}}\label{parameter_B}\\
E&=&{2\over M_{pl}^2}(2\pi)^{-3}\alpha^{\prime-2} \sum_i e^{-{8\pi
K_i\over 3 g_s M_i}} \eeqa
where we have defined
\be
{\cal V}=\ti{\cal V}\int d^6y \sqrt{\ti g} e^{-4\omega(y)}\equiv \ti{\cal V}C \a^{\prime 3} \\
\ee
with $\ti{\cal V}=e^{6 u_0}\equiv X_0^{3/2}$ and so
\beqa
\a'={2\ti{\cal  V} C\over g_s^2 (2\pi)^7M_{pl}^2}.
\eeqa
$c$, $D$, $W_0$ are taken to be the values in \cite{kklt}, except for a minus sign in $W_0$ that cancels
out the effect of
taking $\langle Y \rangle = \pi$: $c=0.1$, $D=1$, $W_0=1/25000$. However, we will make
use of scaling properties of
the potential (\ref{rescaling}), (\ref{rescaling2}) to explore a wider space of parameters.

Up to this point all the analysis has been model independent; the differences among models will depend in the concrete
value of $h_{p-3}$ as well as in the dependence of the warping and the compactification volume on the parameters of the model. We will
analyse this issues in detail in the forthcoming sections.

There are other factors that can change the value of $A$ and $B$ in a rather model independent way: these are the electric and magnetic flux one can turn on
the world volume of the brane, and the external, constant NSNS $B$ field
that can be consistently added to any model without spoiling the GKP/KKLT construction.
We will briefly review these issues in the section 6.

%\subsection{Slow-roll analysis}
%
%When can we expect sufficient number of e-folds? A simple argument is
%as follows. If we assume that the initial condition for the K\"ahler
%modulus sets it close to the minimum and ignore its role in inflation,
%the number of e-folds %for the case when $k=1$
%is given by \be N(T)\approx \frac{AB}{M_{pl}^2} \int_{T_e}^{T_b}
%{V^2\over V'}dT\,, \ee and so $N(T)\propto e^{(5-p)\omega}h_{p-3}$.
%Since generically $e^\omega<<1$, only for $p\geq 5$ can one hope to
%obtain sufficient number of e-folds, provided that $h_{p-3}={\cal
%O}(\a^{\prime (p-3)/2})$. This is what we will assume in the first
%part of the paper. Once one has the initial conditions to obtain
%this, we can investigate if other phenomenological requirements such
%as density perturbations constraints and spectral index constraints
%can be met. For all cases we must be careful not to choose initial
%conditions which will decompactify the compact space. This will
%place constraints on how large the tachyon contribution will be
%compared to the contribution from the K\"ahler modulus.
%the standard slow roll parameter are given by\cite{rae,sv}
%\ba
%\epsilon_1&\approx& {1\over 2 AB}{V'^2\over V^3} \\
%\epsilon_2&\approx& {1\over AB} (-2 {V''\over V^2}+3 {V'^2\over
%  V^3})\,. \label{slowroll}
%\ea
%These parameters turn out to be very small during the inflationary
%phase.

\subsection{The particular case of $D9-\overline{D9}$ pairs}

We show in this section why it is impossible to get inflation with the $D9-\overline{D9}$
system if one does not  provide extra sources of tension to the branes, such as electromagnetic fluxes.
The action for the $D9-\overline{D9}$ system in the absence of these extra sources is
\beqa
S={2\sqrt{2}\over M_{pl}^2}T_9\int d^{10}x\sqrt{g_{MN}+\p_MT\p_NT}.
\eeqa
Using the general metric metric (\ref{metric_gral}) and expanding inside the square root one finds 

\beqa
A\simeq{g_s \over 2\pi^2\a'}{\int d^6y \sqrt{\ti g}~e^{-2\omega}\over \int d^6y \sqrt{\ti g}~e^{-4\omega}},\quad\quad
B\simeq{\int d^6y \sqrt{\ti g}~e^{-4\omega}\over \int d^6y \sqrt{\ti g}~e^{-2\omega}}.
\eeqa
It follows that the number of e-folds is 
\beqa
N\simeq\alpha' AB\simeq{g_s\over \sqrt{2}\pi^2}.
\eeqa

\subsection{Applying the open-string completeness conjecture}

We have three second-order coupled differential equations for $X$,
$T$ and $a$. We need six initial conditions. Now, Sen's open
string completeness conjecture puts a constraint on the form of
the tachyon equation (\ref{tachy}). This equation has been derived
from a four dimensional action, that has in turn been derived from
a ten dimensional effective action where the tachyon field is
invariant under the symmetry $x_0\to -x_0$ as dictated by
conformal field theory analysis. This implies that all the
effective physics regarding the tachyon has to be invariant under
this symmetry. In particular, equation (\ref{tachy}) has to be.
The easiest and most natural way to fulfil this requirement is to
set \beqa
T(x_0)&=&T(-x_0)\nonumber\\
X(x_0)&=&X(-x_0)\label{initial}\\
a(x_0)&=&a(-x_0)\nonumber
\eeqa
Note that these conditions preserve the time translational symmetry $x_0\to x_0\to x_0+h$, since e.g. $T(x_0+h)=T(-x_0-h)$.
This and (\ref{initial}) imply that $\p_t f(t)|_{t=0}=0$, for $t=x_0+h$ and $f=T,u,a$. As we see, Sen's completeness
conjecture is putting
constraints on the initial conditions:

\begin{itemize}
\item{We have the freedom to choose the origin of time, that is,
$h$ above. We set $h=0$.} \item{This automatically implies, by
Sen's conjecture, \beqa \dot a(0)=\dot T(0)=\dot X(0)=0
\label{constraints1} \eeqa} \item{Friedman's equation puts
another constraint in the equations of motion: \beqa H^2&=&{1\over
6} \b \left(\dot X\over 4X\right)^2+{1\over 6}\hat{V}(X)+ {1\over
6}{A\cdot V(T)\over \sqrt{1-B\dot T^2}}\label{hubble}-{k\over
a^2}. \eeqa This has to be valid at all times. However, the time
derivative of this equation is the equation of motion for $a(t)$,
and, since this is a first order differential equation, the
constraint is valid for all times if imposed at $t=0$. Then, we
have the initial condition \beqa {1\over 6}\hat{V}(X(0))+ {1\over
6}A\cdot V(T(0))-{k\over a(0)^2}=0\label{constraint2} \eeqa All
the solutions we are going to consider are such that
 $\hat{V}(X(0)), V(T(0))>0$, and thus we must set $k=1$.
}
\end{itemize}
So we have fixed four initial conditions in (\ref{constraints1}), (\ref{constraint2}) out of the six initial ones. Eq. (\ref{constraint2})
fixes the value of $a(0)$ in terms of $T(0)$ and $X(0)$.

Now, we can fix the value of $X(0)$ demanding that it stays in the value that minimises the potential (in absence of tachyon)
in $t=0$. Given that $X(t)$ is going to be
symmetric with respect to the time and it is going to move all the time around this minimum, we can consider this to be (approximately)
its value at the
beginning of time.
With respect to $T(0)$, it is usually argued that the initial value of the inflaton field
should be close to 1 in Planck units. Assuming this
to be true, and following (\ref{canon}), the initial value of $T$ should be chosen to be of order 0.1.

\subsection{Evolution of the system}

Before we proceed putting numbers in the system, we find it convenient to give an overview of the dynamics of the system.

We want to consider the tachyon to be the only field driving
inflation, whereas the volume modulus field remains as a constant.
The tachyon field evolution is given by equation (\ref{tachy})
whereas its coupling to the metric is given by (\ref{evolutiona}).
We can decouple the modulus field from the system supposing $\dot
X=0$ and $\hat V(X_{min})=0$, in order for the cosmological
constant not to drive inflation. If we manage to get dynamically
$\dot X=0$, then we can use the results of \cite{rae,sv} to
study the tachyon evolution in a slow-rolling regime.

To consider the stabilisation of the field $X$, first we have to be careful that $A$ is sufficiently small for the total potential
at $t=0$, $\hat{V}(X)+A(X)V(T_0)$ has a minimum different from $X=\infty$. This necessary condition can be rephrased in terms of an
approximate inequality
between the Hubble constant during inflation and the gravitino mass
first derived in \cite{lindekallosh}\fn{The authors of
\cite{lindekallosh} expressed their worry about that such a constraint since it seems to imply a either a
high-energy susy breaking scale
along the lines of the phenomenological scenarios depicted in \cite{splitsusy} if one wants to get acceptable $H$, or a unnaturally
low $H$ if one wants to get $m_{3/2}\sim {\rm TeV}$. We argue that what is going to measure the scale of susy breaking is
the magnitude of the soft terms that is observable in the SM throat. As argued in \cite{softerms,CQS} is it perfectly possible to obtain
phenomenologically acceptable soft terms with such a high value for $m_{3/2}$.}

\be
H\lesssim m_{3/2}.
\label{gravitino}
\ee

Once we have chosen $A$ to fulfil this property we must set the initial value of $X$ in the minimum (or near the minimum) of
the total potential $\hat{V}+AV(T_0)$. Note that this minimum will not be exactly the same of the KKLT potential. We discuss the naturalness
of such a choice in section 7. The contribution of the tachyon potential to the total potential for $X$ starts to decrease as time
grows, and $X$ will undergo small damped oscillations (since the minimum starts displacing towards the KKLT minimum). These
oscillations will quickly stop due to the damping and the possible relevance of them to density perturbations are completely negligible
since they occur just at the beginning of inflation.

Inflation happens while the tachyon stays close to the maximum of
its potential. Once the tachyon abandons this maximum inflation
quickly ends. The potential for $X$ decreases from the initial
value to the KKLT potential, and $X$ undergoes a displacement from
its original minimum to the KKLT minimum.

Another interesting issue is the behaviour of the system under a
rescaling of the form (\ref{rescaling}). It should be noted that,
once the condition  (\ref{gravitino}) is fulfilled, one can make
$W_0$ as big as one wants (but being careful to evaluate all the
relevant $\a'$ corrections) since the inflation is driven by $A$
(and the residual cosmological constant left over after the KKLT
mechanism) while $X$ is at the minimum of the potential. However,
the bigger $W_0$ is, the higher the fine tuning one has to make in
considering the initial value of $X$ to get inflation and not
decompactification. For acceptable values of $W_0$, as those
considered along the paper, no fine tuning in the initial position
of $X$ is necessary.

A last remark is in order. In order to trust our effective
description for the fields different from $T$ it is necessary
that, for all times, $m_s\geq m_{KK}\geq H$. This is in principle
non trivial since the value of $m_s,m_{KK}$ will depend on the
warping in a sensitive way. The open string scale in the
inflationary throat in units where $M_{pl}=1$ is roughly given by
\be m_s^2\sim {1 \over B \alpha'}\,. \label{naive} \ee In order to
see this, note that the tachyon mass sets the string scale.
Expanding the tachyon potential to leading order and rescaling the
tachyon field to canonical form leads to the above result. Note
that $B$ depends on warping and hence the string scale in this
instance is suppressed. The Kaluza-Klein open string scale is set
by \be m_{KK}^2\sim {1\over B V_3^{1/3}}\,, \ee where $V_3$ is the
volume of the 3-cycle that the D6-brane wraps around. Since the
warp factor does not depend on the world-volume directions we can
choose $V_3$ to be greater than ${\cal O}(\alpha')$ and make
$m_{KK}<m_s$.
%Further we can check numerically that using the
%parameters specified in the next section in these formulae, the
%inflation scale is smaller than $m_{KK}$.
Note that the closed string scale and the
closed string Kaluza Klein scale involve the averaged warp factor and
are in principle bigger than the above scales. Moreover, the decay of
the D-brane will give rise to massive closed string modes, but the
open-string completeness conjecture allows us to average their effect
by taking into account the initial conditions.

In the numerical analysis performed in the following section we have
found ${\cal O}(m_s)\gg {\cal O}(H)$,
where we have used formula (\ref{naive}) for $m_s$.
This was possible by turning on a world-volume near critical electric field
which reduces the Hubble scale as to be explained in the next section.

It could be argued that in scenarios in which ${\cal O}(m_s)\sim {\cal O}(H)$ a non-negligible density of primordial
black holes could be produced. While, in accordance with the completeness conjecture the dynamics of the
closed string black holes should be captured by the classical evolution
of the field $T$, quantum fluctuations of $T$ could give rise to primordial open string black holes (not necessarily limited
to the beginning of inflation), whose fast decay could leave a trace on the CMB. Such result would necessarily depend on the
exact relation between $m_s$ and $H$ and then its study lays far beyond the scope of this paper.

%\newpage

\section{Addition of electric, magnetic and NS-NS B-fields}
As stressed above, the addition of electric and/or magnetic in the
world volume of the branes, or, alternatively, constant NS-NS
B-field in the bulk, can significantly enhance the
phenomenological features of a given model. Setting them to zero
is too restrictive.  Let us briefly review how these fields can
affect the inflationary scenario. The DBI action takes the general
form:

\beqa S_{T}=-T_p\int_{M_4\times \Sigma} d^{p+1}x \, V(T)
\sqrt{|g_{ab}+ Q_{ab} + \p_a T\p_b T|} \label{action_tachtwo} \eeqa

Where $Q_{ab}= B_{ab} + 2\pi \alpha' F_{ab}$, with $B_{ab}$ the
bulk NS-NS antisymmetric tensor field and $F_{ab}$ is the gauge
field strength.

\subsection{Electric field}

To illustrate the effect of an electric field on the physical parameters
$A$ and $B$, let us consider the simple case when the 6-d metric is flat.
Let us turn on an electric field along an internal direction labelled by
$m$ which is
consistent with the symmetries of the FRW metric. In this case the
$p+1$ dimensional Lagrangian for the non-BPS brane takes the form

\be
A_p V(T)\sqrt{g_4} \sqrt{1-B\dot T^2-E^2}\,.
\ee
The equation of motion for $E$ is given by
\be
\partial_{m}{\sqrt{g_4}A_p V(T)E\over \sqrt{1-B\dot T^2-E^2}}=0\,,
\ee
This tells us that if $g_4$, $A_p$ and $B$ are independent of the internal
coordinate $m$ then a constant $E$ will satisfy this equation. This is
true for example when the 6-d space is a torus. In this case the
4-d effective
Lagrangian can be rewritten as
\be
\hat A V(T) \sqrt{g_4}\sqrt{1-\hat B \dot T^2}\,,
\label{electric}
\ee
where $\hat A=A \sqrt{1-E^2}$ and $\hat B=B/(1-E^2)$. Thus the total
number of e-folds which is proportional to $AB$ would increase. This is
consistent with the fact that tachyon condensation is supposed to slow
down in the presence of a background electric field as shown in
\cite{sen_electric}. Also we see that this can reduce the Hubble scale
keeping the string scale fixed and hence can be useful in adjusting
the scales in the analysis.

For the case of a KS geometry, things are more complicated since the 6-d
metric will introduce a dependence on $m$ generically. As a result a
constant $E$ solution will no longer be valid. One could however
presumably choose $E$ in a coordinate and time-dependent manner that
improves the number of e-folds and it would be interesting to investigate
this issue further.

\subsection{Magnetic field}
If we turn on a constant magnetic field in the internal dimension, this
will effectively change the Lagrangian to
\be
AV(T)\sqrt{1-B \dot T^2} \sqrt{g_{int}+F_{int}}\,,
\ee
as a result it is easy to see that this leads to effectively increasing
$A$ keeping $B$ unchanged. Since the magnetic field has a topological nature, one
must have 2-cycles in the wrapped
region to be able to turn on this field in a consistent way.

\subsection{Constant NS-NS B-field} One could also consider turning on a
background constant $B_{0m}=e$, where $m$ labels an internal
dimension. In this case the Lagrangian of the probe non-BPS brane takes
the form
\be A V(T)\sqrt{1-B \dot T^2-B e^2}\,,
\ee
as a result by tuning
$B e^2$ close to 1, we could reduce $A\rightarrow A\sqrt{1-B e^2}$ while
increasing $B\rightarrow B/(1-B e^2)$. This in principle could lead to
sufficient number of e-folds while satisfying density perturbation and
spectral index constraints.

Thus we see that having a background $B_{0m}$ field
potentially can resolve problems with tachyon inflation. Also note that
since $A$ is reduced, this lowers the Hubble scale while keeping the
string scale fixed. As to be explained in the discussion this can resolve
a potential problem with scales in tachyon driven inflation models.

\subsection{Issues concerning the addition of fluxes}

One could wonder whether the addition of electric or magnetic field can leave some trace (apart from the trace left by
the decay of the brane without fluxes) in the CMB
after tachyon condensation. In the case of the magnetic field the answer is clearly no: magnetic field corresponds to the presence of
lower dimensional non-BPS branes in the world-volume of the corresponding non-BPS state, and they will produce the same kind of
topological defects as the mother non-BPS brane does. The case of the electric field is more complicated since after tachyon condensation a
density of fundamental strings are left over stretching in the same
direction of the original electric field \cite{sen_electric}. This
happens in particular if the original electric field was turned on in
a homotopically non-trivial cycle and is not expected if the cycle is trivial.

Another interesting issue is that of the fine tuning. While in the concrete models we have analysed
a large amount of magnetic flux seems to be needed to obtain phenomenologically acceptable
results (though this situation might not be generic), this cannot be interpreted as fine tuning. However, in the case of
the electric field it turns out that one must choose $E^2\to 1$ in $\hat A$, $\hat B$ in order to make its contribution relevant.
While from the numerical point of view it seems an undeniable fine tuning, a more careful study of the physical problem shows that
a $E$ is not a sensible parameter to describe the electric field, but rather
$\ti f\equiv {E \over \sqrt{1-E^2}}$, since it is the integral of this field (the conjugate momentum of $A_i$) over an sphere at infinity
what yields the electric charge\fn{We thank A. Sen for pointing out this subtlety about the fine tuning to us.}. 
Thus, a value of $E$ very close to 1 should
not be interpreted as fine tuning since this interpretation only relies on a bad choice of 
variables. This interpretation cannot be given for the NSNS field,
in which case a value of this field very close to 1 should be interpreted as fine-tuning.

\section{Example 1: Klebanov-Strassler throat}

Although one can perform a general analysis of the equations
(\ref{tachy})-(\ref{evolutiona}) for an arbitrary CY (assuming that
there always exists some CY such that the values of $A$ and $B$
computed from its geometry satisfy all the phenomenological
requirements), we find it interesting to analyse in detail the
scenario in concrete geometries.

We consider in this section an illustrative example of a warped
region of a Calabi-Yau which corresponds to a deformed conifold.
We will see that it is very difficult to satisfy the  slow-roll
conditions for inflation in this case. In the next section we
consider a more successful example.

The warped deformed conifold solution of Type IIB sting theory\footnote{ We thank X. Chen, J. Raeymakers and S. Trivedi for prompting
us to study this case in detail and for informing us of
unpublished work in which this possibility was analysed, with
negative results regarding inflation. Our analysis here agrees
with their results, though the generalisation to include electric and magnetic flux in the KS throat is still an open problem.}, 
%%FF added the footnote
also known as Klebanov-Strassler throat \cite{KS}, is one of the few known CY metrics.
A detailed study of its geometry has been carried out in \cite{Candelas,ks_geometry}; we refer to these papers to the interested readers.

%\begin{figure}
%\begin{center}
%\epsfig{file=KS.eps,width=8cm,height=7cm}
%\caption{Embedding of the warped deformed conifold/KS throat in a CY compactification. The radial coordinate $r$ has been explicitly displayed.
%Each point of the throat correspond to a $S^3$, whereas the circles around the symmetry axis correspond to the $S^2$. The form of the
%metric
%(8.4) is valid between $r_0$ and $r=R$. The complete form of the metric \cite{KS} is valid from the tip to $r=R$.
%Since then, the conifold is merged with the CY and neither the KS metric or the approximation are valid.}
%\end{center}
%\end{figure}

The KS geometry once embedded in a general CY manifold has two 3-cycles that are dual to each other; the metric and fluxes are a good solution
of Type IIB supergravity provided that $M$ units of RR 3-form flux $F_3$ are
turned on on the $S^3$ three-cycle that is on the tip of the throat and $-K$ units of NSNS $H_3$ flux are turned on in the dual 3-cycle:
\beqa
{1\over (2\pi)^2\alpha'}\int_A F=M; \hspace{2cm}{1\over (2\pi)^2\alpha'}\int_B H=-K
\eeqa
where $A$ is the $S^3$ and $B$ its dual 3-cycle. The complete metric is very well known and analysed, but
for our purposes let us just say that there are two interesting limits in the geometry. The first one is the geometry
'far from the tip of the throat', that is valid for large radial coordinate $r$. In this case the metric has the approximate form
\beqa
ds_{10}^2=e^{6u_0-6u}{r^2\over R^2}d\bar s_4^2+e^{2u}{R^2\over r^2}(dr^2+r^2ds_{T^{1,1}}^2).
\label{metric_far}
\eeqa
where $T^{1,1}$ is a 5-dimensional space with the topology of $S^2\times S^3$. $R$ is a dimensionful quantity given by
\beqa
R^4={27\over 4}\pi g_s MK \a^{\prime 2}.
\eeqa
The other interesting limit is the tip of the throat, that is a
7-dimensional manifold (the volume of the $S^2$ in $T^{1,1}$ has been shrunk to zero size) with metric
\beqa
ds_7^2=e^{2\omega-6u+6u_0}d\bar s_4^2+e^{2u}\ti R^2 ds_{S^3}^2.
\label{metric_tip}
\eeqa
where $\ti R$ and $e^{2\omega}$ are given by
\beqa
\ti R&=&\sqrt{g_s M}~\a^{\prime 1/2}\\
e^{2\omega}&=&e^{-{4\pi K \over 3 g_s M}}.
\eeqa
Let us now discuss several ways of realising our scenario in the KS geometry.

\subsection{D6-branes localised on the tip of the throat}

The metric in which we must embed the brane is (\ref{metric_tip}),
which is of the form (\ref{metric_gral}) for $\ti g_{mn}=e^{2\omega} \ti R^2g_{mn,S^3}$. Now, in this case 
$h_3=2\pi^2\ti R^3e^{3\omega}$.
If we substitute this value for $h_3$ in (\ref{parameter_A}), (\ref{parameter_B}),
we find (after fixing the volume modulus)

\beqa
A&=&{4\sqrt{2}\pi^3e^{4\omega}\a^{\prime-1}g_s^{5/2}\over X_0^{3/4} C}\\
B&=&e^{-2\omega}.
\eeqa
Now, note that \cite{shamit}
\beqa
C=k\left({2\pi\over 3}\right)^3\left({27\pi\over 4} g_s MK\right)^{3/2}
\label{constraint_giant}
\eeqa
for $k \geq1$ some real number.
The number of e-folds becomes (evading again fine tuning of $T_0$, like in previous section)
\beqa
N\simeq \a'AB ={27\over \sqrt{2}}\left({4\over 27\pi}\right)^{3/2}{g_s e^{-4\pi K\over 3 g_s M}\over k K^{3/2}X_0^{3/4}}
\eeqa
that is, at most, a number of order $10^{-2}$. Note that we cannot make this number any better by turning on electric or
magnetic fields since there are
no 1-cycles or 2-cycles in a $S^3$. This situation resembles the giant inflaton case studied in \cite{shamit}.
%%FF added the previous comment

\subsection{Branes wrapping $T^{1,1}$, away from the tip.}
The case of the non-BPS $D8$ admits an interesting variation on the scenario. One can consider wrapping the D8 in a 5 cycle of $T^{1,1}$
with the topology of $S^2\times S^3$ at a value of $r$ different from zero. While this system dynamically tries to go to the
tip of the throat where it annihilates into the vacuum, one can consider the interesting possibility of turning on a magnetic flux
along the directions of the $S^2$. This magnetic field is topological and takes values in the integers over the area of the 2-cycle.
After pullbacking the KS metric in the world volume of the brane in the presence of magnetic flux, it is easy to see that a potential
for the position of the brane is generated, that stabilises it away from the tip where the 2-cycle shrinks to zero.
This magnetic field would then have the dual role of stabilising the brane far from the throat and enhancing the number of e-folds.
Note that the same comments might apply for $D5-\overline{D5}$ pairs wrapped in the $S^2$ away from the tip.
A detailed investigation of this scenario may deserve further study.

%For D8 branes this possibility is particularly interesting since
%the brane would wrapp the full $T^{1,1}$ space which has
%two-cycles where magnetic fluxes can be turned on. We have seen
%that this substantially improves the situation regarding the
%number of e-folds. Furthermore, magnetic fluxes would act to
%prevent the D8 from going to the tip of the throat, stabilizing it
%at some point inside the throat. A detailed investigation of this
%scenario may deserve further study. %%FF added previous lines but not sure if it is actually true

\subsection{Branes along the throat, wrapping the $r$ direction}
We can consider $D6$ or $D7-\overline{D7}$ pairs wrapping the $r$ direction and one of the
2- or 3-cycles available in the KS geometry. For this to make
sense, the complete KS metric should be used.
%There is no known expression %%FF is this true? I thought it was known but messy
%for $h(\tau)$ in an intermediate
%region of the throat and the best one can do is to use the
%approximate metric (\ref{KSm2}) with the perturbative expression
%for $h(r)$ \cite{kklmmt}
However, for the sake of illustration we will use the simplified form of the metric (\ref{metric_far}),
from a infrared cutoff $r_0\ll R$ to $R$, where the throat is supposed to merge with the whole CY.
%\beqa h(r)={R^4\over
%r^4}\left(1+{g_sM\over K}\left({3\over 8\pi}+{3\over 2\pi}\log
%({r\over R})\right)\right). \eeqa

The DBI action for a brane wrapping the $r$ direction after volume stabilisation is
\beqa
S_{DBI}={2 T_p\over M_{pl}^2}\int d^4x \sqrt{-\bar g} ~\Omega_{p-3} X_0^{p-3\over 4}V(T) \int dr~ R^{p-7}r^3\sqrt{1-{R^2\over r^2}\dot T^2},
\eeqa
where $\Omega_{p-3}$ is a numerical factor coming from the angular integration. Considering the case of very big
warping at the tip of the throat, $e^{-2\omega}=e^{{4\pi K\over 3g_sM}}\gg1$, we find (for $R^2\dot T^2/r^2$ small)
\beqa
\int_{r_0}^R dr~R^{p-7}r^3(1-{R^2\over 2r^2}\dot T^2)\simeq{R^{p-3} \over 4}(1-\dot T^2),
\eeqa
where we have assumed $r_0 \ll R$. The values of the $A$ and $B$ parameters is (using (\ref{constraint_giant}))
\beqa
A&=&(2\sqrt{2}){(2\pi)^{7-p}\a^{\prime (7-p)/2}g_s\Omega_{p-3}X_0^{p-9\over 4}R^{p-9}\over 4 k\left({2\pi\over 3}\right)^3}\\
B&=&2.
\eeqa
The factor of $(\sqrt{2})$ appears in the $D7-\overline{D7}$ case. The number of e-folds is given by
\beqa
N=(\sqrt{2}) {27~(2\pi)^{4-p}\over 2k}\Omega_{p-3}X_0^{p-9\over 4} \left({27\pi \over 4}MK\right)^{p-9 \over4}
g_s^{p-5\over4}.
%\d_H^2&=&(4)\cdot 3^9{(2\pi)^{12-2p}\over 300\pi^2}{g_s^4\over 8k^3}\Omega_{p-3}^2X_0^{p-3\over 2}
%\left({27\pi\over 4}MK\right)^{p-12\over2}.
\eeqa
Note that this number is again something of order $10^{-1}$, what makes it difficult to
reconcile with the cosmological observations.

\section{Example 2: D6 branes in $T^3$ fibres}

As a second model, consider the case of wrapping a non-BPS D6
brane in a 3-cycle of the CY with the topology of a torus. These
cycles are quite generic, since following the work of Strominger,
Yau and Zaslow \cite{SYZ}, every Calabi-Yau that has a mirror is
expected to be a  $T^3$ fibration over some basis. Let us see how
we can combine this topological statement with the GKP setup in
order to perform a quantitative analysis of inflation in such a
setting.

\begin{figure}
\begin{center}
\epsfig{file=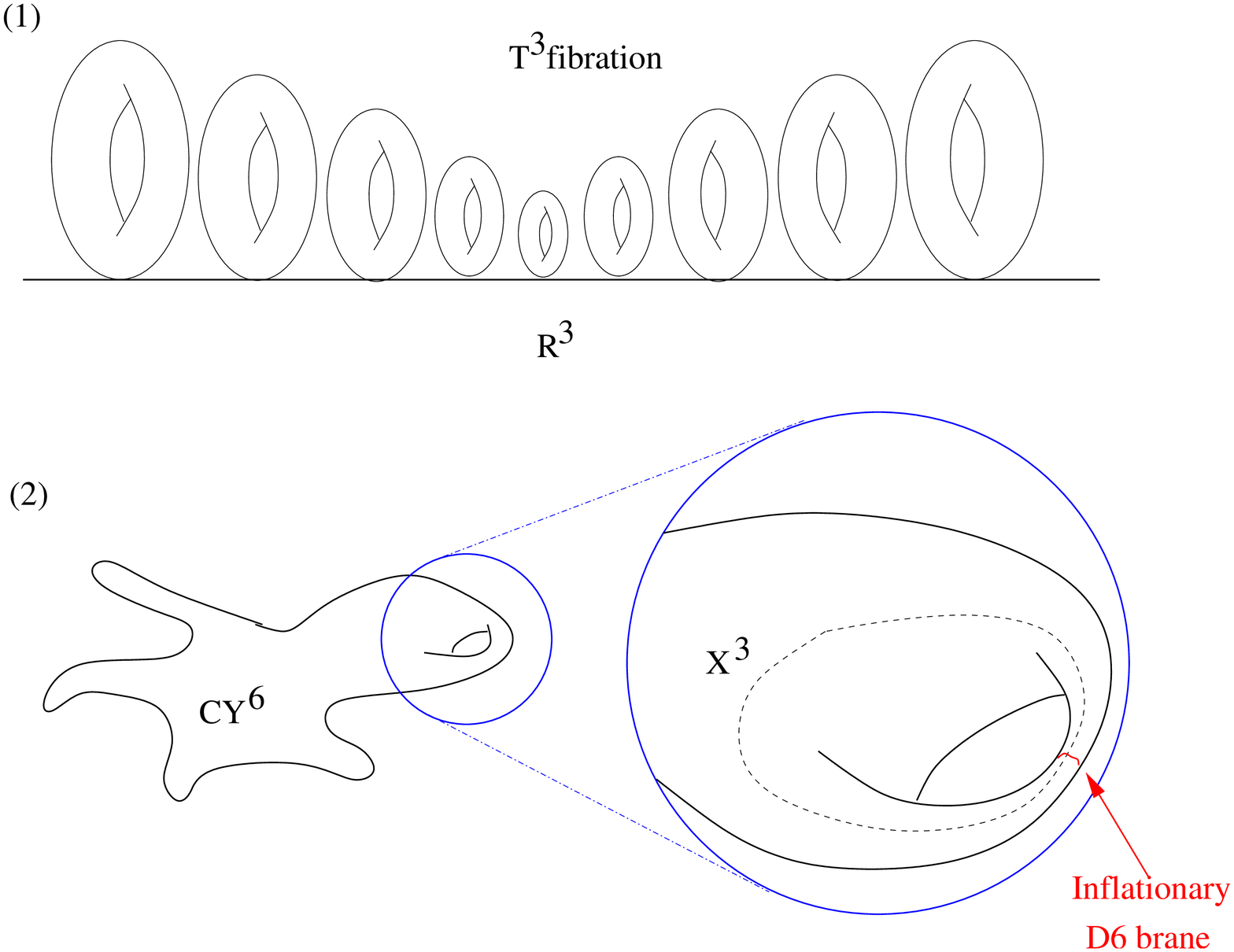,width=10cm} \caption{Depiction of the
CY as a $T^3$ fibration. (1) Locally, the CY can be seen as a
$T^3$ fibration over ${\bf R}^3$. (2) The global structure of the
CY can be very complicated but one can envisage that locally one
can have D6 branes wrapped in the $T^3$ fibre at the point of
maximum warping. Here $X^3$ represents the (3 dimensional) local
base space.}
\end{center}
\label{figfibration}
\end{figure}

Our aim then is to consider in the GKP setup a construction that
is locally ${\bf R}^3\times \ti T^3$, with $\ti T^3$ is a warped
torus (warping in the directions of ${\bf R}^3$). The metric in
the vicinity of that region is of the form \beqa
ds^2=Z^{-1/2}d\bar s_4^2+ e^{2u_0}Z^{1/2}({\rm d}\vec{\bf
x}^2+{\rm d}\vec{\bf y}^2) \eeqa where $\vec{\bf x}$ are the
coordinates in the ${\bf R}^3$ patch and $\vec{\bf y}$ are the
coordinates in the $T^3$ fibre. Let us assume that $Z$ does not
depend on $y$. The equation of motion for the warping function
$Z(x)\equiv e^{-4\omega}$ reads \beqa (\ti\nabla^2Z){\rm vol}=g_s
F_3\w H_3 \label{warpingtorus} \eeqa with ${\rm vol}=\sum dx_i\w
dy_i$, the volume form of the (unwarped) internal manifold, and
the $\ti \nabla^2$ refers to the Laplacian in the internal
unwarped coordinates. We can wrap the RR $F_3$ flux along the
$T^3$ directions and the NSNS $H_3$ flux along the dual directions
($X_3$ in figure 7). Dirac quantisation conditions imply roughly
that \beqa
F_3&=&{M\over V_{T^3}}\\
H_3&=&{K\over V_{X^3}}. \eeqa with $K$, $M$ integers. ISD
condition for the fluxes will imply that \beqa {M\over
V_{T^3}}={K\over V_{X^3}} \eeqa with $V_X^3$, $V_{T^3}$ the
corresponding volumes in $\alpha'$ units, and these numbers $M$,
$K$ have to satisfy a tadpole condition of the form \beqa
MK\lesssim 10^4. \eeqa The solution of (\ref{warpingtorus}) will
be of the form \beqa e^{-2\omega(x)}=Z^{1/2}(x)\simeq
g_s^{1/2}\sqrt{{M\over V_{T^3}}{K\over V_{X^3}}}|x|.
\label{solution_t} \eeqa There will be in principle other
dependences in $x$ as well as integration constants; one should
remark at this point that (\ref{warpingtorus}) is not the complete
equation but it has some contribution from point-like charges like
O3s and D3s that we will consider to be far apart from our
solution. The complete solution will be unique and will fix all
the arbitrariness of (\ref{solution_t}). Let us assume that $x\sim
1$, $V_{X^3}\sim 1$, $V_{T^3}\sim 1$. This will imply, after
moduli stabilisation \beqa B=e^{-2\omega}=g_s^{1/2}(KM)^{1/2}.
\eeqa Now, $A$ is given in the D6 case by \beqa
A={g_s\sqrt{2}(2\pi)^{7-p}e^{(7-p)\omega}X_0^{p-9\over 4}\a^{\prime
{(1-p)\over 2}}\over C}h_{p-3}={2\pi\sqrt{2} g_s^{3/4} \over \alpha'
k(KM)^{3/4} X_0^{3/2}}, \eeqa where we have used the fact that
$C=\int e^{-4\omega}\sqrt{\ti g}\sim k(KM)$ for $k$ some positive
real number greater than 1, $h_3=\alpha^{\prime
3/2}V_{T^3}/X_0^{3/4}$ and we have chosen $V_{T^3}\sim 1$. It is
quite clear that one will not get a sufficient number of e-folds
just will these $A$ and $B$. The addition of $n$ units of magnetic
flux will increase $A$ by a factor of $n/$Area, where Area is the
area in $\a'$ units of the $T^2$ inside $T^3$ where we turn on
this magnetic field. We can consider this area to be of order 1. In addition, we can consider
a value for the electric field $f\equiv {1\over \sqrt{1-E^2}}$ as a time-varying Wilson line along one 
of the 1-cycles of the $T^3$. After all this
we get 
\beqa A={2\sqrt{2}\pi ng_s^{3/4} \over \alpha'f k(KM)^{5/4}
X_0^{3/2}}, \quad B=f^2 g_s^{1/2}(KM)^{1/2}\eeqa 
Can we satisfy the constraints for the number of
e-folds and the density perturbations for some values of these
parameters? The number of e-folds in absence of fine-tuning for
the initial value of the tachyon field is roughly given by \beqa
N\simeq \a' AB\simeq 10^2 \label{n_e-folds} \eeqa and the density
perturbations are given by \beqa \d_H^2={1\over 25 \pi^2}{H^4\over
p+\rho}={(2\pi)^7\over 300\pi^2}{\a^{\prime 2}g_s^2\over X_0^{3/2}
C~ {\rm sinh}^2({T_0\over \sqrt{2\a'}})}A^2B. \label{den-city}
\eeqa
%what implies (in absence of fine tuning for $T_0$)
%\beqa
%{A^2B g_s^2\over X_0^{3/2} C}\simeq 10^{-12}.
%\eeqa
The phenomenological constraints (number of e-folds and density
perturbations) amount to have, in terms of the parameters of the
model \beqa
N\simeq \a'AB\simeq{2\sqrt{2}\pi n fg_s^{5/4}\over k(KM)^{3/4}X_0^{3/2} }&\simeq& 10^2\\
\d_H^2\simeq {(2\pi)^9\over 300\pi^2}{2g_s^4 n^2\over k^3(KM)^3
X_0^{9/2}}&\simeq&10^{-10} \eeqa both constraints can be easily
satisfied with standard values for the parameters like $g_s\sim 0.1$, $M\sim K\sim 10$, $k X_0^{3/2}\sim
10^{3}$, $n\sim 10^{3}$, $f\sim 10^4$.  
This implies a large volume of the
compactification manifold ($\sim 10^{6}-10^7$ in units of $\a'$),
consistently with having all the sources far away from the
toroidal construction, and also a somewhat large electric and magnetic flux. 
Many other combinations of these parameters may also give the desired
values. 

%for instance, $g_s\sim 10^{-2}, KM\sim 10^4$, $k
%X_0^{3/2}\sim 10^5$, $n\sim 10^{11}$, etc. Notice that turning on
%NS-NS backgrounds also contribute to improve
%the number of e-folds and give many more possible combinations of
%parameters that give rise to inflation. %%FF added these comments

\section{Results}
\subsection{Inflation by non-BPS $D6$ brane}
Here we present numerical results for inflation generated by a non-BPS
D6 brane, in order to show explicitly how the dynamical evolution of the system proceeds and 
to illustrate graphically issues like moduli stabilisation, the end of inflation and the big crunch-big bang transition. 
It is more convenient to work in units where $M_{pl}=1$. We
will assume that the numbers for $A$ and $B$ can be adjusted by
turning on a combination of bulk fluxes and world-volume fluxes. The initial conditions and values for various parameters are
as follows:
\ba
E&=& 3.9\times 10^{-12}\nonumber \\
A&=& 2\times 10^{-3} X^{-9/4}\,, \quad B= 5\times 10^{9} \left({X\over
  X_{min}}\right)^{3/2}\nonumber \\
c&=&0.1\,, \quad ~~~~\,D=1 \nonumber \\
 \quad W_0&=&{1\over
  25000}\,,\quad T(x^0=0)=0.1\,,
\ea
with the rescaling parameters
$$
\mu=800,\quad \lambda=1,
$$
so that the effective{\footnote{Note that in the units
    of \cite{DouglasKachru} where $l^2=\sqrt{2\pi} \alpha'=1$, $W_0\approx 0.1$.}}
$W_0\approx 0.9$. We have chosen $E$ and $\Lambda$ so that
the minimum for $\hat{V}(X)$ is of ${\cal O}(10^{-16})$ at $X_{min}=124.636$. This
is because we do not want inflation to be driven by the cosmological
constant term\fn{The value we have chosen for $\Lambda$ is $5.14 \times 10^{-11}$.}
We have also conducted separate numerical tests with a
more fine-tuned $E$ such that the minimum is even lower and have
obtained very similar results. Including
the contribution from the tachyon, the effective potential takes the
shape of the thinner line in figure 2.
\begin{figure}[h]
\begin{center}
\epsfig{file=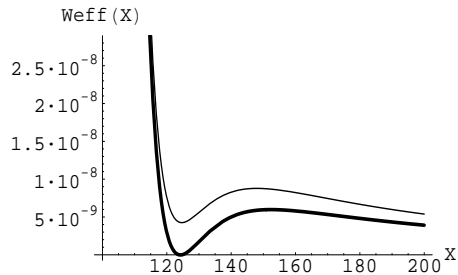,width=6cm, height=4cm}
\caption{Effective potential with and without tachyon
  contributions. The thicker line shows the potential at the end of
  tachyon condensation while the thinner is the potential at the beginning.}
\end{center}
\end{figure}
Note that the initial potential has a minimum as well. This property
is crucial to prevent decompactification from happening.
The behaviour for the number of e-folds and $a(t)$ as
a function of time are exhibited in figure 3. Using the parameters
specified above, we get around 90 e-folds.
\begin{figure}[h]
\begin{center}
\epsfig{file=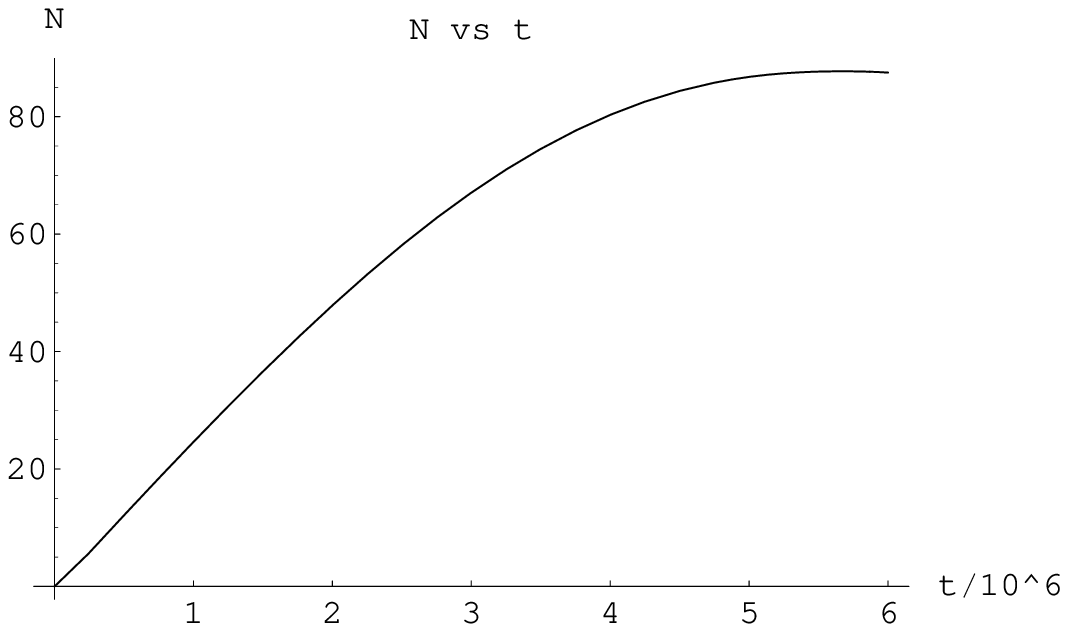,width=5.5cm, height=3.7cm}\hskip 2cm
\epsfig{file=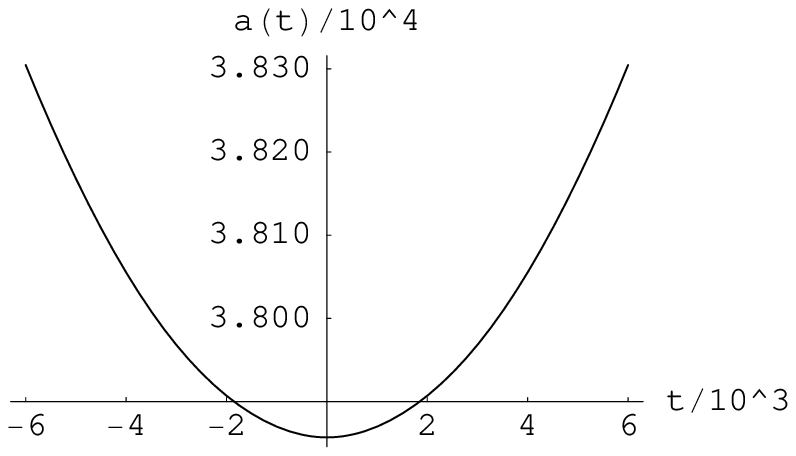,width=5.5cm, height=3.7cm}
\caption{Number of e-folds vs $t$ and $a(t)$ vs $t$.}
\end{center}
\end{figure}
We have displayed the time reversal property of $a(t)$ explicitly.
Other important quantities such as the volume of the compact manifold,
tachyon potential, $\rho+3p$ are demonstrated below. We will not
display $t<0$ explicitly in these graphs although it should be assumed
that the solution is time symmetric.
\begin{center}
\FIGURE[h]{\epsfig{file=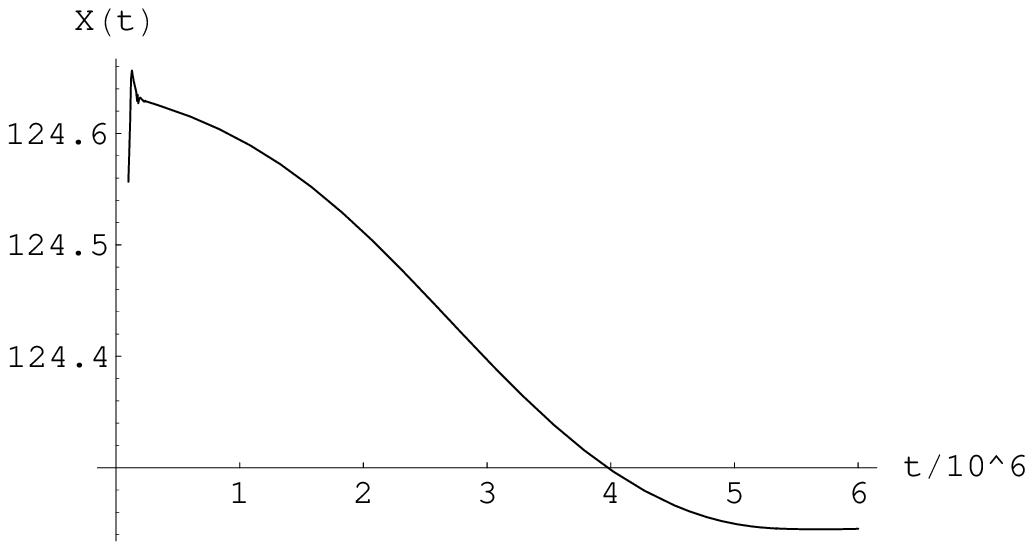,width=6cm, height=4cm}\hskip 2cm
\epsfig{file=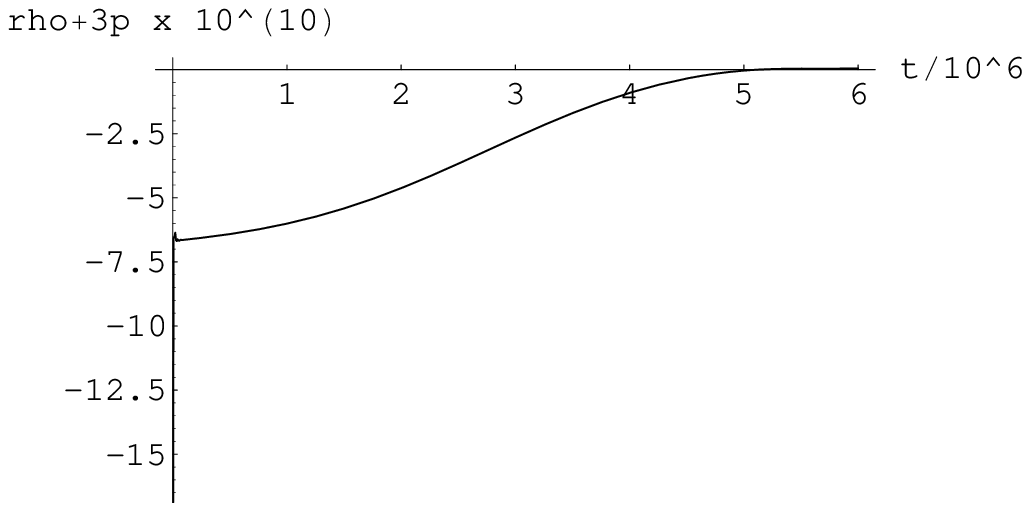,width=6cm, height=4cm}
\caption{$X(t)$ vs t and $\rho+3p$ vs $t$.}
}
\end{center}

\begin{figure}[h]
\begin{center}
\epsfig{file=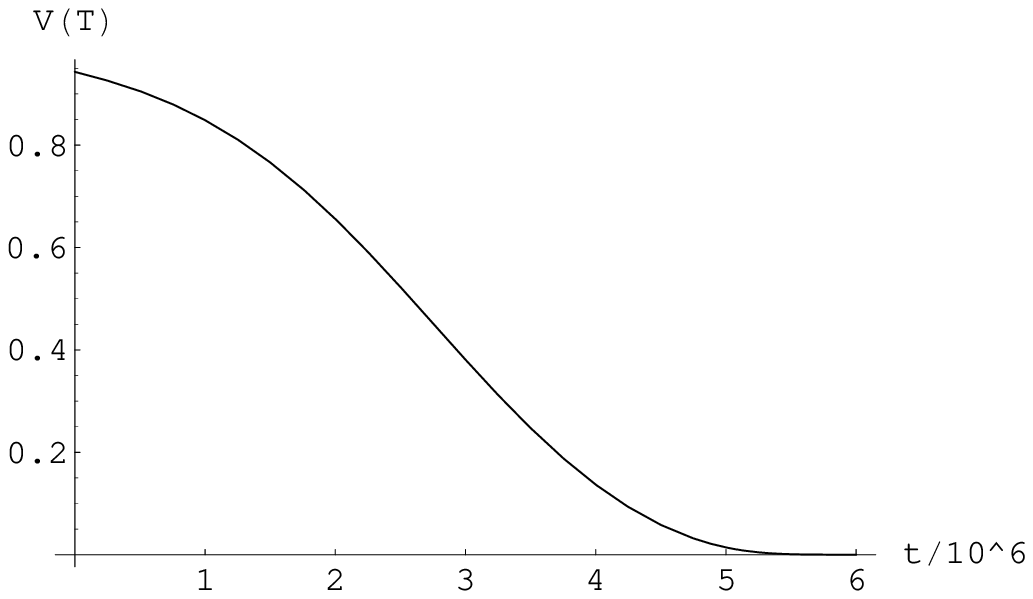,width=6cm, height=4cm}
\caption{$V(T)$ vs $t$.}
\end{center}
\end{figure}
\noindent Finally, we have plotted the deviation of the spectral index from
unity as a function of the
number of e-folds. This has the behaviour given in figure 6.
\begin{figure}[h]
\begin{center}
\epsfig{file=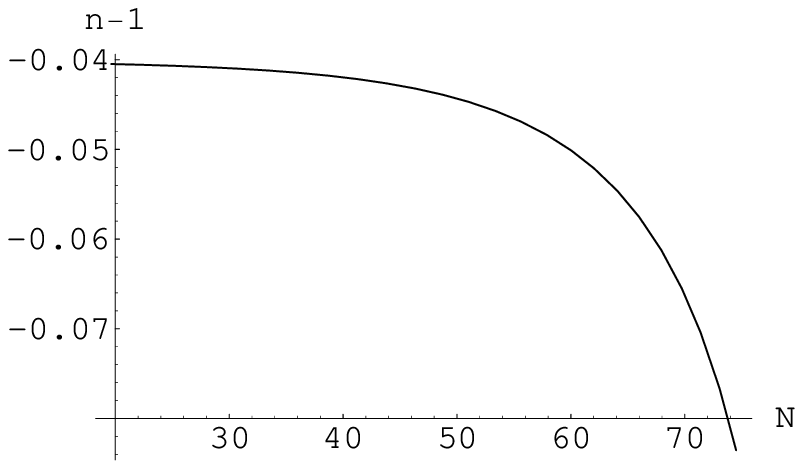,width=6cm, height=4cm}
\caption{Plot of $n-1$ vs $N$}
\end{center}
\end{figure}

We have used the formulae
%\fn{We have made an educated guess for the
%  formula of ${\cal P}(N)$ for the case of the tachyon action. This formula
%reduces to the usual one in the case of a single, canonically normalized field %or for the tachyon field in the slow roll regime.
%It takes into account, however, all possible powers of $\dot T$ and takes into %account even all possible derivative corrections,
%that will surely play a role in creating inhomogeneties outside the slow-roll r%egime.} \cite{racetrack}
\ba
{\cal P}(N)&=& {1\over 25 \pi^2} {H^4 \over p+\rho} \\
n-1&=& {d \ln {\cal P}(N) \over d N} \\
\delta_H &=& \sqrt{{\cal P}(N)}|_{N=30}\,. \label{dperturb}
\ea
This suggests that the spectral index is between $0.95$ and
  $0.96$. However, this value can be pushed closer to unity by making
  $AB$ larger. We have also estimated the density perturbation
and found it to be $\approx 2\times 10^{-5}$ at $N\approx 30$. We
choose $N=30$ since the COBE normalisation has to be imposed at
approximately 60 e-foldings before the end of inflation. Since the
total number of e-foldings is 90 in this example, we calculate the
spectral index and the density perturbation when $N=30$.
One can perhaps tune the parameters to make these results
better. In this model, the energy scale during inflation is $H\approx
10^{14}$ GeV. Also the amplitude of gravitational waves produced
during inflation is $h\approx {H\over M_p}\lesssim 10^{-5}$ which
satisfies observational constraints on CMB anisotropy.
%In this example
%involving the D6, there is an additional peculiarity. Noting that
%\ba
%A & \sim & {2\over (2\pi)^{5/2}} {g_s^{7/4}\over \sqrt{2}}\, e^\omega\\
%B &\sim & e^{-2 \omega}\,,
%\ea
%we see that the size of the compact manifold essentially drops out of
%all the physical quantities apart from $\alpha'$ in
%$V(T)$. Furthermore, note that in this special case, the density
%perturbation $\propto \sqrt{A^2 B}$ is independent of the warping and
%hence the warping parameter can be used to get good results for the
%number of e-folds and the spectral index. In general, the larger the
%number of e-folds, the closer the spectral index is to unity. However,
%one has to be careful that the Hubble scale is lower than the
%Kaluza-Klein scale.
One can potentially get the same physics as in our
example in models with large extra dimensions as in \cite{CQS}. This is very interesting and merits further investigation.

\subsection{Fine-tuning in our models}

The usual drawback of inflationary models is the huge amount of
fine tuning required to make the model work. We have argued that
our model suffers less from fine tuning than other inflationary
models, particularly stringy ones. The reasons are two-fold. First
the warp factor dependence in the non-BPS D-brane action. This
warp factor depends on the fluxes and therefore can take
relatively large values. This includes the electric and magnetic
fluxes\footnote{The NS-NS field can also help for inflation
although making the quantity $1-B_{ab}^2$ small is actually a fine
tuning, which is different from the electric field case.}. The
explicit models that we have require large volume and large 3-form
and magnetic fluxes. We then exchange fine-tuning by large fluxes.
Notice that in the successful cases of $T^3$ fibrations, the
warping of the metric induced by the fluxes is not exponential. %%FFF the previous line
The second reason is that the tachyon field is not subject to the
standard $\eta$ problem of supergravity potentials since its 4d
action is not manifestly supersymmetric.

We can still check how much fine-tuning is needed for the parameters
of our model. We have numerically verified that small variations of
the values we have chosen give rise to small variations in the number
of e-folding, and as expected any reduction of the number of
e-foldings can be compensated by the choice of the warp factors.

%\section{$D7-\overline{D7}$ pairs on a 4-cycle of an arbitrary CY}

%\section{The case of D9 branes}

\section{Discussion and Outlook}

Let us finish
with a discussion of some relevant issues of our scenario:

\begin{itemize}

%%FF modified title
\item{{\bf General bounds on scales in any tachyon driven
inflation model}

We saw that the string scale is given by ${c\over B\alpha'}$ where $c$ is
an unknown constant. The Hubble scale is given by ${A V\over 6}$.
In order for the string scale to be bigger than the Hubble scale we need

\be {c\over B\alpha'}>{A V\over 6}\,. \ee Using the formula for
the number of efolds and denoting the integral $\int_{T_e}^{T_b}
{V^2\over V'}dT=\Delta/\a'$, we have \be 6c>{N V\over 2 \Delta}\,.
\ee Thus for $c\sim O(1)$ we find that $\Delta$ has be quite large
which is only possible if the tachyon was sitting very close to
the top of the potential and if $AB\sim O(1)$. However, this
immediately leads to the problem that $\eta\propto {1\over AB}\sim
O(1)$ and thus we are not in the slow-roll regime any more.
Although the number of e-folds can be quite large, the density
perturbation constraints will no longer be satisfied.  There are
two possible resolutions to this impasse. Firstly, $c$ could be
larger. With $c$ of $O(10)$ the numerical example in the paper
would yield a string scale twice as large as the Hubble scale.
Secondly, one could turn on a world-volume electric field which
will reduce the Hubble scale keeping the string scale fixed. This
in principle will make the scales work in the right direction.
Notice that these constraints  allow a large number of e-folds but
not arbitrarily large.} %%FF slightly modified previous paragraph

%\vspace{1cm}

%\newpage
\item{{\bf Time reversal symmetry of the solutions}

All our cosmological solutions are invariant under a change $t \to -t$. This kind of solution arise from
the evolution of the open-string field $T$ that describes (following Sen's conjecture) the decay of the non-BPS object
only if the initial conditions allow for this time-reversal symmetry of the system. This does not imply that more general,
time asymmetric solutions cannot exist: it implies that such solutions cannot be described in such a simple way, with the tachyonic action
\ref{actioneg1}.

Time reversal symmetry implies that if the universe is to end up in a big crunch singularity for $t=t_0$, then it must have
started from a big bang singularity in $t=-t_0$. Analogously, if the fate of the Universe is to expand forever at $t\to \infty$,
it is because it started from an infinitely extended initial state in $t\to -\infty$.
In this sense, we must clarify what we understand by big bang. In order to get a proper understanding of the situation, we
should identify the big bang to be the state of the universe at $t=0$.

Note that in our model the universe can never be singular at $t=0$,
since following equation (\ref{constraint2}) $a(0)>0$. This
(non singular) big bang is always a consequence of a previous (non
singular) big crunch, note that
the null energy condition of
\cite{turokseiberg} that forbade big crunch-big bang transition does not apply here since here $k=1$.}
\item{{\bf Exit from inflation}

The fact that we are able to track the system for all time relies
heavily on
Sen's conjecture. By studying the (classical) evolution
of a open string field we are actually considering all the complicated
interactions of the set of closed string fields produced during
the decay. It is the conjecture what allows us to trust the effective
theory for all
times\fn{But only with respect to the D-brane decay,
and only if we interpret the results as quantum expectation values.
We can only trust the validity of the whole effective action
(including the rest of the fields) if $H\leq m_s$ for all times.}
and what allows us to put numbers in the end of inflation.

With respect to a mechanism for reheating, we assume that a mechanism
like that proposed in \cite{bbc} applies. The SM is supposed to
be placed in the tip of a throat whose warping is much bigger than
that of
the inflationary throat (see figure 1). In this case,
the graviton wave-function is peaked in the SM throat and the closed
string
modes produced during the decay can serve (enhanced by the
gravitational blue-shift) to excite the SM degrees of freedom.

In any case, a complete study of the problem of reheating is beyond the scope of this paper.}

\item{{\bf (De)compactification and Big Crunch singularity}

The evolution of the system presents the following peculiarity.
If the initial ($t=0$) value of $X$ is such that its potential
energy is much bigger than the KKLT barrier, during its evolution it
will overshoot
KKLT barrier and the universe will undergo
decompactification. If this decompactification is fast enough, then

\beqa
{\ddot a\over a}\simeq -\left(\dot X\over 4X \right)^2
\eeqa

very soon and the four dimensional
space is led to a big crunch singularity. In these cases, The usual damping term for the
scalar field given by $H$ is very small or even negative, thus not disturbing (or even enhancing for $H<0$) the
decompactification process.
On the other hand, if the $t=0$ conditions are
such that $X$ is trapped
inside the KKLT well, it will always be there
(since its evolution suffers from a damping term) and stabilises
very soon. Then it is a good approximation to consider it
as a constant for all times, while the four dimensional universe inflates.

We have however found some marginal cases
in which the position of $X$ is tuned so that it
overshoots the barrier with very little kinetic energy and thus its slow-rolling
evolution turns on inflation of the four dimensional
universe, producing at the same time decompactification. However, the amount
of inflation obtained in this way is always very small
compared to the case with the volume modulus stabilised.
Moreover, these cases are very
marginal and one has to fine tune the position of $X$ to get such a behaviour.

Another case that leads to decompactification and big crunch are those with a very
big initial value of $T$. In this case, tachyon decay proceeds very fast and the closed string moduli starts to oscillate.
These oscillations are negligibly damped and, after some time, get enhanced by a negative value of $H$, what asymptotically leads
to decompactification plus big crunch.

We find, then, in a broad majority of cases, either inflation with
moduli stabilised or decompactification with big crunch.

We are tempted to conclude that, since such KKLT potentials
(with a set of metastable vacua and then a runaway behaviour for
an infinite vev of the moduli fields) are supposed to be generic
 in string theory, such a behaviour of change of roles between
compact and non-compact dimensions under cosmological evolution
is quite generic, independently of the model of inflation considered.}
This seems to agree with recent studies
\cite{rob}.

%\item{{\bf Initial conditions and naturalness}
%
%In a model with a time reversal property, the initial conditions ($t=0$) are actually sent to $t\to -\infty$. This fact, and the
%relationship between big crunch and decompactification (or expansion and compactification) imply that the concept of naturalness
%of the initial value of $X$ (see \cite{racetrack}) must be revised for models of the kind considered here. If we consider
%natural to have inflation, then the internal dimensions must be compact. This implies that for $t=-\infty$ they must be compact,
%and this imply that they will never decompactify during inflation.}
\item{{\bf $\alpha'$-corrections}}

Although we took the first
  $\alpha'$-correction for the superpotential into account, there are
  also potential corrections to the tachyon effective action. Not
  much is known about these. Suppose there were terms in the effective
  action of the form $((\partial_\mu \partial^\mu)^n T)^{2k}$. Since
  the contraction of the indexes is with $e^{-2\omega}/V g^{\mu\nu}$, it is
  quite possible that there is an enhancement compared to the unwarped
  case (where these terms are volume suppressed) unless the higher
  tachyonic derivatives are vanishingly small. Although this is true
  during slow-roll, this is not necessarily true during the period of
  exit from inflation as our graphs show. Thus these terms, if present, could
  become very important during this phase. Taking them into account in
  a systematic manner is beyond the scope of this paper. However,
  these terms will not alter the conclusion that sufficient number of
  e-folds can be produced with the tachyon as the
  inflaton. Furthermore, we have checked numerically that in our
  models, $e^{-2\omega}\ddot T/V<<1$ as a consequence of which we believe that
  our results may correctly capture the physics even towards the end
  of inflation.

\end{itemize}

In summary, we have presented a successful  inflationary scenario in
string theory with  moduli stabilization. We  made progress in
obtaining inflation in a natural way by controlling the amount of
fine tuning in terms of the flux-induced warp factors. It is worth
mentioning that the exponentially warped regions such as the
Klebanov-Strassler geometry turned out to be less suitable to
satisfy the slow-roll conditions. Whereas regions with no
exponential warping, such as the $T^3$ fibrations, can give rise
to inflation, requiring large volume and fluxes. Therefore we
exchange the fine tuning of parameters needed in other string
inflation models by the introduction  of large fluxes. %%FF added previous lines to make things clear.
The slow-roll conditions are not satisfied by $D9- {\overline{D9}}$ systems
unless magnetic fluxes could be included. They are also difficult to
satisfy for $D6$ branes on KS throats. $D8$ and $D5- {\overline{D5}}$ may
satisfy them once magnetic fluxes are included. The most concrete
successful example we provided was for $D6$ branes on  $T^3$
fibrations. We also expect similar conditions for 
$D7- {\overline{D7}}$ systems wrapping on  four-cycles 
with non-trivial two-cycles inside. A detailed study of all these configurations
lies outside the scope of this article.

Our scenario also opens new avenues to explore string
cosmology in flux compactifications. The implications of the
open-string completeness conjecture have proved very efficient to
extend the model to earlier regions of the universe, leaving
interesting possibilities for speculation about the initial
conditions.

There are many open questions remaining that would need further
attention.
 We have made several simplifying assumptions to keep
the model as simple as possible. In particular we have assumed that
the
complex structure moduli and the dilaton have been stabilised and only
concentrated on the K\"ahler structure modulus/ tachyon system.
In the same way that the tachyon dependence  alters the potential
for $X$ during the brane decay time, it may also affect the other
moduli.
In both cases there is a dependence on these fields of the tachyon
 part of the potential.
The dilaton dependence
can be seen by tracing the $g_s$ factors of $V(T)$ in the total scalar
potential. The complex structure moduli associated to the size of the
 three-cycle that the D6 brane is wrapping, would also appear in the
 tachyon potential.
It would be interesting to  follow the cosmological evolution of
all these fields also but this is beyond the scope of this
article. We may argue that in the absence of the non-BPS brane the
dilaton is fixed at weak coupling, and since its potential
blows-up at infinity, then the coupling to the tachyon gives rise
to only a runaway dependence (recall that the dilaton field $\tau$
is such that Im($\tau)\propto 1/g_s$) that may change only
slightly the cosmological evolution of $\tau$. We hope to address
this issue in the future.

Also, recently \cite{bbcq,CQS}, an extension of the KKLT scenario
has been put forward such that the potential stabilises at  an exponentially
large volume, with all other moduli also fixed. The coupling of these
systems, that require at least two complex structure moduli, to the
tachyon may lead to interesting cosmological implications. The
exponentially large volume is welcome in order to better trust the
effective field theory description.

Finally, we have compared our scenario with the brane-antibrane
and racetrack inflation, especially regarding the advantages on fine tuning and
initial conditions. It may also be possible to consider different stages of
inflation  in which several of these mechanisms may be at work at
different times in the evolution of the universe.

To conclude, some words about the physical interpretation of time-reversible situations.
Clearly, the time symmetry of this scenario asks for a proper interpretation
of the full cosmological scenario.
At the moment we can say that starting from $t=0$, the evolution of the
universe is just like in standard eternal topological inflation, driven
in this case by the decay of a non-BPS brane. We may
consider if there is any physical meaning at all regarding the
$t<0$ region of our spacetime. Notice that the initial condition $\dot T
=0$ is different from the finite velocity assumed in the S-brane
interpretation of the tachyon potential \cite{sbrane}. Unlike that case, our scenario is completely time-symmetric:
in a sense, there is no preferred arrow of time. The case $v=0$ of  \cite{sbrane} would correspond to $T(0)=0$ in our case
(that is a classically static situation),
while our general case has not any analog along the lines of  \cite{sbrane} (note that for $T(0)\neq0$ the tachyon never reaches the
top of the potential). It would be
interesting to develop a full cosmological picture based on our results.

\acknowledgments{We thank P. Berglund, C.P. Burgess, P. G. C\'amara, 
X. Chen, J. Conlon, S. Hartnoll, L. Ib\'a\~nez, P. Kumar, F. Marchesano, L. Martucci,
  J. Raeymakers, P. Silva, K. Suruliz, D. Tong and S. Trivedi and especially A. Sen and A. Uranga for very
  useful discussions. We thank A. de la Macorra for early
  collaboration on related subjects.  DC is supported by the University of
  Cambridge. FQ is partially supported by PPARC and a Royal Society
  Wolfson award.
AS is supported by a postdoctoral fellowship
  from PPARC and a research fellowship from Gonville and Caius
  College, Cambridge.}
%\newpage

\end{document}